\documentclass[11pt]{article}
\usepackage[a4paper,margin=2.3cm]{geometry}
\usepackage{amsmath,amssymb,amsfonts}
\usepackage{physics}
\usepackage{braket}
\usepackage{graphicx}
\usepackage[dvipsnames]{xcolor}
\usepackage{tikz}
\usetikzlibrary{decorations.markings,decorations.pathmorphing}
\usepackage[super,sort&compress,comma]{natbib}
\usepackage{setspace}
\usepackage{authblk}
\usepackage[font=small,labelfont=bf,labelsep=period]{caption}
\usepackage{booktabs}
\usepackage[hidelinks]{hyperref}

\definecolor{docnotelinkcolor}{rgb}{0,0,0} 

\newif\ifshowflags \showflagsfalse

\renewcommand{\figurename}{Fig.}
\newcommand{\Fig}[1]{Fig.~\ref{#1}}
\title{\textbf{Twist-tunable clapping modes and their tunnelling signatures in bilayer cuprates}}
\author[1]{Silvia Neri}
\author[2]{Marcel Franz}
\author[1]{Demetrio Vilardi}
\affil[1]{Max Planck Institute for Solid State Research, Heisenbergstra\ss e 1, 70569 Stuttgart, Germany}
\affil[2]{Department of Physics and Astronomy and Quantum Matter Institute, University of British Columbia, Vancouver, British Columbia V6T 1Z4, Canada}
\date{\small Correspondence: Silvia Neri, s.neri@fkf.mpg.de}

\begin{document}
\maketitle

\begin{abstract}
\noindent Collective modes of the superconducting order parameter encode its internal structure, but tunnelling spectroscopy offers no clear way to identify them, as bosonic features such as dip–hump structures can also arise from phonons, magnetic resonances or the band structure. Twisted bilayer cuprates near a $45^\circ$ twist realize a chiral, time-reversal-symmetry-breaking state hosting generalized clapping modes. Computing the tunnelling density of states across this phase, we find that the relative-phase mode and a hybrid amplitude mode remain sharp in-gap excitations pinned just below the pair-breaking edge. The two modes redistribute spectral weight in opposite ways, so their effects partly compensate in the momentum-integrated spectrum, whereas the amplitude mode dominates along the nodal direction. Because the mode energies follow the twist-tunable minigap, the resulting resonances move about three times faster than the minigap itself, and the outer resonance shifts opposite to any feature from an extrinsic boson. This provides a twist-angle test for intrinsic collective modes.
\end{abstract}


\section*{Introduction}
The collective modes of a superconductor encode the symmetry of its order parameter. Every superconductor carries an amplitude (Higgs) mode, associated with fluctuations of the magnitude of the order parameter~\cite{PekkerVarma2015,ShimanoTsuji2020,BarlasVarma2013,Matsunaga2013,Matsunaga2014,Measson2014,PhysRevLett.120.117001}. When the pairing state is built from several components, the collective-mode spectrum becomes richer. In multiband superconductors, an out-of-phase oscillation of the phases of two condensates gives rise to the Leggett mode~\cite{Leggett1966,PhysRevLett.99.227002}. In superconductors with competing pairing channels, Bardasis--Schrieffer modes appear as excitonic pair oscillations in a channel that is energetically close to, but distinct from, the one that condenses~\cite{BardasisSchrieffer1961,PhysRevB.92.094506,PhysRevX.4.041046}. In time-reversal-symmetry-breaking (TRSB) superconductors, such as $s+id$ or $d+id'$ states, oscillations of the relative phase and of the relative amplitude of the two components give rise to generalized clapping modes~\cite{Balatsky_2000,SaulsWuChung2015,Poniatowski2022}. These are in-gap modes whose frequencies depend on the ratio between the two condensed components, so resolving them would give direct spectroscopic access to the internal structure of an unconventional pairing state.

Tunnelling spectroscopy offers a possible route to their detection~\cite{Bardeen1961,TersoffHamann1983,Fischer2007}. Coupling to a sharp bosonic mode enters the quasiparticle self-energy and imprints a characteristic dip--hump structure on the single-particle density of states (DOS)~\cite{Eliashberg1960,ScalapinoSchriefferWilkins1966}. Different bosons leave their fingerprints in this way, from phonons in conventional superconductors~\cite{McMillanRowell1965} to spin excitations in the cuprates~\cite{Chubukov2003,PhysRevLett.87.067005,Zasadzinski2006,PhysRevLett.85.3261,PhysRevLett.83.1652}. These fingerprints are, however, hard to interpret: a magnetic resonance, a particular phonon, or structure in the bare electronic DOS can each produce a similar feature~\cite{PhysRevLett.87.067005,Zasadzinski2006,Devereaux2004,Shan2012,Levy2007,Lee2006}.

The idea that tunnelling can probe pair fluctuations goes back to Ferrell and Scalapino, who considered a metal above its superconducting transition~\cite{Ferrell1969,PhysRevLett.24.1052}, and Gherghetta and Nambu treated the amplitude and phase modes as bosons exchanged by the quasiparticles and computed the resulting self-energy corrections~\cite{PhysRevB.49.740}. The idea has recently been revisited for unconventional superconductors~\cite{LeeSteiner2023}, and two experiments have reported DOS features attributed to superconducting collective modes: a resonance in single-layer NbSe$_2$, interpreted as a Leggett mode~\cite{Wan2022NbSe2}, and a peak--dip--hump feature in the kagome superconductor CsV$_{3-x}$Ta$_x$Sb$_5$, interpreted as either a Leggett or a Bardasis--Schrieffer mode~\cite{Hu2024}.

In neither case does the lineshape alone fix the assignment, and both works leave open whether the feature stems from a genuine collective mode or from another source of the same qualitative structure. What is missing is a control parameter that moves the intrinsic mode while leaving extrinsic bosons essentially unchanged.

Twisted bilayer cuprates provide such a parameter. Stacking two layers of Bi$_2$Sr$_2$CaCu$_2$O$_{8+\delta}$ (Bi2212), each a $d$-wave superconductor, with a controlled relative twist $\theta$ realizes a two-component order parameter. In a window of twist angles around $45^\circ$ the bilayer is predicted to enter a chiral TRSB state: the two components lock at a relative phase near $\pi/2$, and the two time-reversed configurations $\pm\pi/2$ become distinct, degenerate ground states~\cite{Can,PhysRevB.105.064501,PhysRevB.109.045111,Song2022,Volkov2025}. Signatures consistent with this state have been reported in twisted Bi2212 Josephson junctions~\cite{Zhao2023}. Other junction experiments report either a Josephson coupling that persists at all twist angles~\cite{Zhu2021} or a strong suppression of the critical current near $45^\circ$~\cite{Martini2023}, so the existence and extent of the chiral phase remain under active investigation. Within this phase the condensate supports generalized clapping modes~\cite{Poniatowski2022,Balatsky_2000,SaulsWuChung2015}, whose frequencies are set by the relative weight of the two components and can therefore be tuned with the twist (\Fig{fig:concept}).

Here we compute the signatures of these modes in the tunnelling DOS across the TRSB phase. The relative-phase mode and one of the two amplitude modes remain sharp in-gap excitations, while the other amplitude mode is pushed into the quasiparticle continuum. This allows us to compare, on equal footing, the effect of an amplitude and of a phase mode on the quasiparticles. Both imprint dip--hump features that shift with the mode energy, but they redistribute spectral weight in opposite ways, so their contributions to the momentum-integrated DOS partly compensate. The angle-resolved spectrum is dominated by the amplitude mode along the nodal direction. Finally, we show that because both modes are pinned to the pair-breaking edge, the twist dependence of the resonance energies, plotted against the minigap measured in the same spectrum, separates an intrinsic collective mode from extrinsic bosons without requiring any knowledge of the coupling strength or of the precise twist angle.

\section*{Results}
\subsection*{Chiral superconductivity in twisted bilayer cuprates}
We describe tunnelling into the top layer by an effective low-energy model obtained by integrating out the bottom layer~\cite{PhysRevLett.132.017002} (Methods). The resulting single-layer Bogoliubov--de Gennes (BdG) Hamiltonian has the off-diagonal element
\begin{equation}
  \Delta_{\mathbf{k}} = \Delta_{1,\mathbf{k}} + e^{i\varphi}\,\Delta_{2,\mathbf{k}},
  \label{eq:OP_intro}
\end{equation}
where $\Delta_{1,\mathbf{k}}$ is the real intrinsic $d$-wave component of the top layer and $\Delta_{2,\mathbf{k}}$ is the real proximity-induced component inherited from the rotated bottom layer, with amplitude $\Delta_2\sim t^2/\Delta_1$ set by the interlayer tunnelling $t$. The interlayer relative phase $\varphi$ is controlled by $\theta$. Solving the coupled nonlinear gap equations~\cite{PhysRevB.53.3598} gives an effective order parameter of the chiral form (Methods and Supplementary Note~1)
\begin{equation}
  \Delta_{\mathbf{k}} =\Delta_0\big[ (1 + r\cos 2\theta \cos \varphi)\,f_{d_{x^2-y^2}} \pm i\,r\sin 2\theta\,\sin \varphi\, f_{d_{xy}}\big]\equiv\Delta_0\chi^\pm_{\mathbf{k}},
  \label{eq:OP_final}
\end{equation}
where $f_{d_{x^2-y^2}}=\cos 2\phi_{\mathbf k}$ and $f_{d_{xy}}=\sin 2\phi_{\mathbf k}$, $\phi_{\mathbf k}$ is the Fermi-surface angle measured from a top-layer antinode, $r=\Delta_2/\Delta_1$, $\Delta_0\equiv\Delta_1$, and the two signs label time-reversed states. The relative phase follows from a Ginzburg--Landau (GL) analysis~\cite{Can},
\begin{equation}
  \varphi_{\min}=\arccos\!\left( \frac{\cos 2\theta}{4\mathcal{K}} \right),
  \label{eq:GL}
\end{equation}
valid for $|\cos2\theta|<4\mathcal K$, where $\mathcal K$ is the ratio of the second- to first-harmonic interlayer Josephson energies (Methods). \Fig{fig:phasediagram}a compares equation~\eqref{eq:GL} with a direct minimization of the microscopic free energy; the fit gives $\mathcal{K}=0.136$, which we use throughout to map $\theta$ onto $\varphi$. In the TRSB window $\theta_c^-<\theta<\theta_c^+$ the spectrum is fully gapped, with a minimum gap $\Delta_{\min}(\varphi)=\Delta_0 r\sin2\theta\sin\varphi$ along the top-layer nodal direction and a maximum gap $\Delta_{\max}(\varphi)$ along the antinodal one (\Fig{fig:concept}b).

\subsection*{Generalized clapping modes across the chiral phase}
In the TRSB phase it is convenient to work with the chiral form factors $\chi^\pm_{\mathbf k}$ of equation~\eqref{eq:OP_final}, which satisfy $\chi^-_{\mathbf{k}}=(\chi^+_{\mathbf{k}})^*$~\cite{Poniatowski2022}. Choosing the $+$ state, the equilibrium gap is $\Delta_0\chi^+_{\mathbf{k}}$, and the uncondensed channel $\chi^-_{\mathbf{k}}$ encodes the relative-amplitude ($a$) and relative-phase ($b$) fluctuations. Together with the Higgs fluctuation $h$ of the condensed channel, these form the Gaussian fluctuation kernel derived in Methods and Supplementary Note~2. The collective excitations separate into a phase sector, containing a single relative-phase mode with propagator $D_b$, and an amplitude sector, containing two modes $A^\pm$ that arise from the hybridization of the Higgs and relative-amplitude fluctuations.

The relative-phase mode is the analogue of the Leggett mode in multiband superconductors~\cite{Leggett1966}: it carries no net charge and is therefore not lifted to the plasma frequency by long-range Coulomb interactions (but see Discussion for the role of interlayer charging). Whether a mode is a sharp excitation is decided by its position relative to the pair-breaking threshold. With $\Delta_{\min}(\varphi)$ and $\Delta_{\max}(\varphi)$ the extrema of $|\Delta_{\mathbf{k}}|$ on the Fermi surface, the two-particle continuum at zero momentum transfer and $T=0$ begins at $2\Delta_{\min}(\varphi)$. A mode below this threshold cannot decay into quasiparticle pairs; a mode within the continuum is damped. The scale $2\Delta_{\max}$ marks the largest local pair-breaking gap rather than an upper edge of the continuum.

\Fig{fig:modes} shows the three spectral functions obtained from the full Gaussian kernels, before the pole approximation used for the tunnelling calculation below. The three modes behave very differently. The relative-phase mode (\Fig{fig:modes}a,d) tracks the pair-breaking threshold $2\Delta_{\min}(\varphi)$ and sits just below it over a window of $\varphi$ around $\pi/2$, where it appears as a single sharp peak. The same behaviour is observed for $A^+$ (\Fig{fig:modes}c,f), while $A^-$ (\Fig{fig:modes}b,e) lies inside the continuum and shows only a broad, damped hump. We therefore focus on the relative-phase mode and $A^+$, the only coherent excitations in the relevant parameter regime. At $\varphi=\pi/2$ their energies are $\Omega_b=39.9$~meV and $\Omega_{A^+}=38.6$~meV, bound by $5$--$6$~meV below $2\Delta_{\min}=45$~meV; towards the edges of the TRSB window the binding energy shrinks rapidly (\Fig{fig:discriminant}b).

\subsection*{Amplitude and phase modes act oppositely on quasiparticles}
The coupling of quasiparticles to mode $i$ enters the Nambu self-energy through a vertex $\hat\Gamma_i$.
With $\tau_{x,y,z}$ the Pauli matrices in Nambu space and $\tau_\pm=(\tau_x\pm i\tau_y)/2$, the vertices read
\begin{equation}
\begin{aligned}
    \hat\Gamma_{h}(\phi)&=\tau_+\chi^+ +\tau_-\chi^-
    =\operatorname{Re}\chi^+\,\tau_x-\operatorname{Im}\chi^+\,\tau_y,\\
    \hat\Gamma_{a}(\phi)&=\tau_+\chi^- +\tau_-\chi^+
    =\operatorname{Re}\chi^+\,\tau_x+\operatorname{Im}\chi^+\,\tau_y,\\
    \hat\Gamma_{b}(\phi)&=i\tau_+\chi^- -i\tau_-\chi^+
    =\operatorname{Im}\chi^+\,\tau_x-\operatorname{Re}\chi^+\,\tau_y ,
\end{aligned}
\label{eq:vertices}
\end{equation}
for the Higgs, relative-amplitude and relative-phase fluctuations. At each point of the Fermi surface the three vertices are vectors of the same length $|\chi^+(\phi)|$ in the transverse $(\tau_x,\tau_y)$ pseudospin plane (\Fig{fig:did_asy}b). The Higgs and relative-amplitude vertices are mirror images of each other under $\tau_y\to-\tau_y$, that is, under time reversal, while the relative-phase vertex is always orthogonal to the relative-amplitude one.

How a vertex acts on the quasiparticles depends on its orientation relative to the local gap $\chi^+(\phi)$. Decomposing the $\tau_+$ component of each vertex along $\chi^+$ and $i\chi^+$ (\Fig{fig:did_asy}c),
\begin{equation}
\hat\Gamma_h\to|\chi^+|\,(1,\,0),\quad
\hat\Gamma_a\to|\chi^+|\,\big(\cos 2\alpha,\,-\sin 2\alpha\big),\quad
\hat\Gamma_b\to|\chi^+|\,\big(\sin 2\alpha,\,\cos 2\alpha\big),
\label{eq:vertexproj}
\end{equation}
with $\alpha(\phi)=\arg\chi^+(\phi)$. The component along $\chi^+$ modulates the magnitude of the local gap, while the component along $i\chi^+$ only rotates its phase. The Higgs vertex is parallel to the gap at every $\phi$. The relative-amplitude and relative-phase vertices instead rotate with the phase of the gap: at $\phi=0$, where $\chi^+$ is real, $\hat\Gamma_a$ is parallel to the gap and $\hat\Gamma_b$ perpendicular to it, while at intermediate angles both have a finite component along $\chi^+$. The sharp amplitude mode entering the self-energy is the hybrid
\begin{equation}
    \hat\Gamma_{A^+}(\phi)=\cos\beta_{A^+}\,\hat\Gamma_{h}(\phi)+\sin\beta_{A^+}\,\hat\Gamma_{a}(\phi),
    \label{eq:GammaAplus}
\end{equation}
where $(\cos\beta_{A^+},\sin\beta_{A^+})$ is the normalized eigenvector of the amplitude kernel at the pole $\Omega_{A^+}(\varphi)$. At $\varphi=\pi/2$ we find $\cos\beta_{A^+}\simeq0.50$ and $\sin\beta_{A^+}\simeq-0.87$, so the $A^+$ vertex is predominantly of relative-amplitude character.

The bare DOS $N_{\mathrm{bare}}(\omega)$ [dotted in \Fig{fig:did_asy}d] has two features: a shoulder at the proximity-induced minigap $\Delta_{\min}$ and a coherence peak at $\Delta_{\max}$, the intrinsic Bi2212 gap. Coupling to a boson mirrors these singularities in the self-energy, shifted by the mode energy $\Omega_i$, producing dip--hump features at $\Delta_{\min}+\Omega_i$ and $\Delta_{\max}+\Omega_i$ and their particle--hole partners at negative bias (Methods). When only $A^+$ is coupled (\Fig{fig:did_asy}d), spectral weight is redistributed towards the gap edges: the coherence peaks at $\pm\Delta_{\max}$ are enhanced and weight is added inside the gap. The relative-phase mode does the opposite (\Fig{fig:did_asy}e): weight is removed from the coherence peaks and moved towards the minigap shoulders at $\pm\Delta_{\min}$, so that the gap edges are smeared rather than sharpened.

This opposite behaviour reflects the orientation of the vertices relative to the local gap. A vertex component along $\chi^+$ modulates the magnitude of the gap and sharpens the gap edges; a component along $i\chi^+$ rotates its phase and smears them. Because $\hat\Gamma_a\perp\hat\Gamma_b$ at every $\phi$, the pure relative-amplitude and relative-phase modes would have exactly opposite effects. The amplitude mode entering the self-energy, however, carries a Higgs component that is everywhere parallel to the gap, so $\hat\Gamma_{A^+}$ is not orthogonal to $\hat\Gamma_b$ and the compensation is only partial. With both modes included (\Fig{fig:did_asy}f), the enhancement caused by $A^+$ is largely offset by the suppression caused by the relative-phase mode, and the low-energy spectrum stays close to the bare one. The resonances on the tails survive, because they sit at slightly different energies, $\Delta_{\min,\max}+\Omega_{A^+}$ and $\Delta_{\min,\max}+\Omega_b$.

\subsection*{Tunnelling spectra across the chiral phase}
\Fig{fig:waterfall_int} shows the momentum-integrated DOS with both modes included, for relative phases between $\varphi=\pi/6$ and $\pi/2$, that is, twist angles between $31^\circ$ and $45^\circ$. The spectrum is symmetric about $\varphi=\pi/2$, since $\theta\to90^\circ-\theta$ maps $\varphi_{\min}\to\pi-\varphi_{\min}$ and leaves $\Delta_{\mathbf k}$ unchanged; plotting $\varphi>\pi/2$ therefore adds no new information.

\Fig{fig:waterfall_int} shows the momentum-integrated DOS with both modes included, for relative phases between $\varphi=\pi/6$ and $\pi/2$, that is, twist angles between $31^\circ$ and $45^\circ$. As $\varphi$ is reduced, the minigap closes and the spectrum evolves from the fully gapped, flat-bottomed shape at $\varphi=\pi/2$ to an almost V-shaped one at $\varphi=\pi/6$, while the coherence peaks move slightly outwards. The two mode energies stay close to each other over the whole range, so the pink and blue markers nearly coincide and the features of the two modes cannot be resolved separately. The resonance at $\Omega_i+\Delta_{\min}$ disperses most strongly with $\varphi$: it lies on the tail outside the coherence peak at $\varphi=\pi/2$, crosses the peak around $\varphi\simeq\pi/4$, and ends up inside the gap at $\varphi=\pi/6$. The resonance at $\Omega_i+\Delta_{\max}$ stays on the tail and shifts more weakly. In the momentum-integrated spectrum all these features are weak, appearing as kinks on the tails of the coherence peaks.

The angle-resolved DOS is much more sensitive to the coupling, because the resonance is not averaged over the Fermi surface and sits at the local energy $\Omega_i+\Delta(\phi)$. \Fig{fig:waterfall_k} shows $N(\phi,\omega)$ along the antinodal ($\phi=0$) and nodal ($\phi=\pi/4$) directions. Along $\phi=0$ (\Fig{fig:waterfall_k}a) the spectrum is BCS-like, with sharp coherence peaks at $\pm\Delta(0)$; the mode coupling produces a step at $-(\Omega_i+\Delta(0))$ and a small hump at the corresponding positive energy, both of which shift towards the coherence peak and weaken as $\varphi$ is reduced. Here the coupling is dominated by the relative-phase mode, but the two contributions act in opposite directions and largely compensate, so that only weak features survive. Along $\phi=\pi/4$ (\Fig{fig:waterfall_k}b) the spectrum is set by the minigap, and the resonance at $\pm(\Omega_i+\Delta_{\min})$ appears as a well-defined dip. This nodal spectrum is set almost entirely by $A^+$, whose vertex couples most strongly to quasiparticles near the gap minimum; the relative-phase contribution is negligible there. Momentum-resolved tunnelling, as provided by the quantum twisting microscope~\cite{Inbar2023}, would therefore isolate the amplitude mode.

\subsection*{A twist-angle test for intrinsic collective modes}
Although the features in \Fig{fig:waterfall_int} are weak, their \emph{positions} carry a clear fingerprint of their origin. Because both sharp modes are bound states of the condensate, their energies are pinned just below the pair-breaking edge, $\Omega_i(\theta)\lesssim 2\Delta_{\min}(\theta)$ (\Fig{fig:discriminant}a). The inner resonance therefore sits at $\Delta_{\min}+\Omega_i\simeq 3\Delta_{\min}$ and moves roughly three times faster than the minigap. By contrast, a boson that is extrinsic to the chiral order parameter---a phonon, or a mode tied to the intrinsic Bi2212 gap such as the spin resonance---produces a feature at $\Delta_{\min}+\Omega_0$, which can move no faster than $\Delta_{\min}$ itself. This motivates plotting the feature energy against the minigap $\Delta_{\min}$ read off from the shoulder of the \emph{same} spectrum, rather than against the nominal twist angle. The resulting plot is self-calibrating: it does not depend on the mapping of equation~\eqref{eq:GL}, on the interlayer tunnelling $t$, on the coupling strength, or on how accurately the twist angle is known, and it is insensitive to spatial variations of the local gap~\cite{Pan2001}, since $\Delta_{\min}$ and the resonance are measured at the same position.

\Fig{fig:discriminant}c,d compares the two scenarios over the window $31^\circ\le\theta\le45^\circ$. We take the extrinsic boson energy to coincide with $\Omega_b$ at $\theta=45^\circ$, the least favourable case. For the inner feature, the collective modes give a slope $d(\Delta_{\min}+\Omega_i)/d\Delta_{\min}\simeq 2.5$, compared with $1$ for a fixed-energy boson and $0.3$ for a boson that scales with $\Delta_{\max}$. The outer feature is even more discriminating, because the slope changes \emph{sign}. As the twist moves away from $45^\circ$, $\Delta_{\max}$ grows while $\Omega_i$ collapses, so $\Delta_{\max}+\Omega_i$ \emph{decreases} (slope $+0.8$ against $\Delta_{\min}$), whereas any feature at $\Delta_{\max}+\Omega_0$ with $\Omega_0$ independent of, or increasing with, $\Delta_{\max}$ \emph{increases} (slopes $-0.7$ and $-1.5$). At $\theta\approx31^\circ$ the intrinsic and extrinsic inner features are separated by $\approx20$~meV, an order of magnitude more than the broadening $\Gamma_0$. The conclusions are unchanged for $t=20$ and $15$~meV (Supplementary Note~4), where the collective-mode slopes approach the limiting values $3$ (inner) and $2+d\Delta_{\max}/d\Delta_{\min}>0$ (outer).

Two caveats set the experimental requirements. First, $\Delta_{\min}$ is stationary in $\theta$ near $45^\circ$, so the inner feature moves by less than $3$~meV between $45^\circ$ and $40^\circ$: a convincing measurement requires devices spanning roughly $31^\circ$--$40^\circ$. Second, the bound states lie within $\Gamma_0$ of the pair-breaking edge for $\theta\lesssim37^\circ$ (relative-phase mode) and $\theta\lesssim34^\circ$ (amplitude mode) (\Fig{fig:discriminant}b). The slope test thus separates features intrinsic to the chiral order parameter from extrinsic bosons across the whole window, whereas identifying a \emph{bound} collective mode, as distinct from the onset of the two-particle continuum, requires resolving the offset $2\Delta_{\min}-\Omega_i$, which reaches $5$--$6$~meV near $45^\circ$.

\section*{Discussion}
We have studied how the collective modes of a TRSB superconductor imprint on tunnelling spectra, using twisted bilayer cuprates as a platform in which the relative phase between the two pairing components, and with it the mode energies, can be tuned. Both in-gap modes---the relative-phase mode and the hybrid amplitude mode $A^+$---produce dip--hump features at $\Delta_{\min}+\Omega_i$ and $\Delta_{\max}+\Omega_i$ that track the mode energies as the twist is varied.

The clapping modes of a TRSB state offer an unusual opportunity to compare an amplitude and a phase mode on equal footing. Both are sharp, in-gap excitations with nearly equal energies, so their effects on the quasiparticles differ only through the vertices. Because the relative-amplitude and relative-phase vertices rotate with the phase of the gap and are orthogonal at every point of the Fermi surface, the amplitude mode sharpens the coherence peaks while the phase mode smears them. In the momentum-integrated DOS these effects largely compensate; the compensation is only partial because the hybrid $A^+$ also contains a Higgs component. The angle-resolved DOS is more informative: along the nodal direction it is dominated by $A^+$ and shows well-defined dips that follow the mode energy, while along the antinodal direction the coupling is dominated by the relative-phase mode but the two contributions largely cancel. Most importantly, because both modes are pinned to the pair-breaking edge, plotting the resonance energies against the measured minigap distinguishes an intrinsic collective mode from extrinsic bosons through the slope of the inner resonance and the sign of the shift of the outer one.

Our model is deliberately simple: a two-dimensional, particle--hole-symmetric BCS framework at $T=0$ with a phenomenological, small-$\mathbf q$ coupling to the bosons, and several of its approximations deserve comment.
First, and most importantly, the relative-phase fluctuation is generated here by the pair susceptibility of the effective top-layer model alone. In the bilayer, $\varphi$ is the phase difference between the layers, so its oscillation drives an interlayer supercurrent and hence an interlayer charge imbalance. The associated charging energy, absent from our Gaussian action, acts in series with the electronic compressibility and stiffens the relative-phase dynamics, as for the plasma mode of a Josephson junction~\cite{PhysRevLett.132.017002}. Taking into account that the inertia of the relative-phase mode is set by the quantum capacitances of the two layers in series with the interlayer geometric capacitance, rather than by the quantum capacitance of the top layer alone, the relative-phase mode enters the pair continuum even for $t=30$~meV, the case most favorable to a bound state (Supplementary Note~5). The amplitude mode $A^+$, which to leading order does not change the interlayer phase difference, is much less affected, and the twist-angle test of \Fig{fig:discriminant}, which relies only on the pinning of the resonances to the pair-breaking edge, survives.
We have nevertheless kept both modes as an illustration of a general mechanism rather than as a prediction for the bilayer.
The opposite action of amplitude- and phase-type fluctuations on the quasiparticles, and the resulting partial compensation, follow from the geometry of the vertices, and apply to any chiral condensate that hosts both clapping modes as sharp excitations.
Within the pole approximation, a damped relative-phase mode simply drops out of the self-energy.
For the bilayer we therefore turn to the more realistic parameters of Ref.~\cite{PhysRevB.105.064501}, $\Delta_1=45$~meV and $t=15$~meV, for which the relative-phase mode lies at the continuum threshold even without charging (Supplementary Fig.~2f) and only $A^+$ is retained.
\Fig{fig:t15_combined} shows the resulting spectra.
The proximity-induced minigap is now small, $\Delta_{\min}\le 5$~meV, and the features in the momentum-integrated DOS are weak (\Fig{fig:t15_combined}a).
Along the antinodal direction the spectrum is essentially unchanged by the coupling (\Fig{fig:t15_combined}b), whereas along the nodal direction $A^+$ produces a dip at $-(\Omega_{A^+}+\Delta_{\min})$ that follows the minigap (\Fig{fig:t15_combined}c).
The same feature can be extracted from the momentum-integrated spectrum through its second derivative (Supplementary Note~6).
The twist-angle test relies only on the pinning of $A^+$ to the pair-breaking edge and gives an inner slope of $2.9$ for these parameters (Supplementary Table~1).
Second, the chiral order parameter of equation~\eqref{eq:OP_final} is obtained from symmetry constraints on the gap equations rather than from a fully self-consistent bilayer solution; it coincides with the global-phase form only at $\theta=\pi/4$, $\varphi=\pi/2$ (Supplementary Note~1). Third, the integration over the bottom layer is perturbative in $t/\Delta_1$, whereas the parameters of Ref.~\cite{Can} give $t/\Delta_1=0.75$ and $\Delta_2/\Delta_1\simeq0.56$; for the parameters of \Fig{fig:t15_combined}, $t/\Delta_1=0.33$ and the expansion is better controlled. The visibility of the modes also depends on $t$: for smaller $t$ the TRSB window narrows and the relative-phase mode moves to the continuum threshold (Supplementary Note~3), while $A^+$ remains just below it.

On the experimental side, the predicted features require tunnelling into the CuO$_2$ plane adjacent to the twisted interface, which favours thin top flakes, and twist angles spanning roughly $31^\circ$--$40^\circ$. The spatial inhomogeneity of the Bi2212 gap~\cite{Pan2001} is mitigated by the self-calibrating analysis of \Fig{fig:discriminant}c,d, which uses only quantities measured in the same local spectrum. Momentum-resolved tunnelling with a quantum twisting microscope~\cite{Inbar2023} would further give access to the nodal spectrum dominated by the amplitude mode. Finally, because the chiral phase itself is still debated~\cite{Zhao2023,Zhu2021,Martini2023}, observing resonances that follow the minigap with the predicted slopes would provide independent spectroscopic evidence for the time-reversal-symmetry-breaking state. Our results should therefore be read as the qualitative signatures that chiral collective modes would leave in tunnelling, and as a study of the properties of generalized clapping modes.

\section*{Methods}
\subsection*{Effective model for the top layer}
Integrating out the bottom-layer electrons~\cite{PhysRevLett.132.017002} gives the renormalized single-layer BdG Hamiltonian
\begin{equation}
\mathcal{H}_{\mathrm{eff}}=\sum_{\mathbf{k}} \psi_{\mathbf{k}}^{\dagger}
\begin{pmatrix}
\tilde{\xi}_{\mathbf{k}} & \Delta_{1,\mathbf{k}}+e^{i \varphi} \Delta_{2,\mathbf{k}} \\
\Delta_{1,\mathbf{k}}+e^{-i \varphi} \Delta_{2,\mathbf{k}} & -\tilde{\xi}_{\mathbf{k}}
\end{pmatrix}
\psi_{\mathbf{k}},
\label{eq:Heff_main}
\end{equation}
where $\psi_{\mathbf{k}}=(c_{\mathbf{k} \uparrow}, c_{-\mathbf{k} \downarrow}^{\dagger})^T$ is the top-layer Nambu spinor and $\tilde{\xi}_{\mathbf{k}}$ the renormalized band energy measured from the chemical potential. We write $\Delta_{a,\mathbf{k}}=\Delta_a f_{\mathbf{k},a}$ for $a=1,2$ with the real form factors
\begin{equation}
    f_{\mathbf{k},1}=\cos(2\phi_{\mathbf{k}}),
    \qquad
    f_{\mathbf{k},2}=\cos(2\phi_{\mathbf{k}}-2\theta).
    \label{eq:formfactors}
\end{equation}
The intrinsic top-layer amplitude $\Delta_1$ is unchanged by the integration, and the induced amplitude is $\Delta_{2} \sim t^2/\Delta_1$ (Supplementary Note~1). The reduction is justified near $\theta = 45^\circ$, where the gap nodes of one layer align with the antinodes of the other, so that the bottom layer acts on the top layer only through a proximity-induced pairing term. The positive quasiparticle energy is $E_{\mathbf{k}}=(\tilde{\xi}_{\mathbf{k}}^2+|\Delta_{1,\mathbf{k}}+e^{i\varphi}\Delta_{2,\mathbf{k}}|^2)^{1/2}$, with BdG branches at $\pm E_{\mathbf{k}}$. We use $\hbar=k_B=1$. Minimizing the free energy with respect to $\Delta_{1,2}$ and $\varphi$ yields a TRSB phase over a finite range of twist angles, consistent with Ref.~\cite{Can}. Imposing the constraints of the coupled nonlinear gap equations~\cite{PhysRevB.53.3598} leads to the order parameter of equation~\eqref{eq:OP_final} (Supplementary Note~1).

\subsection*{Ginzburg--Landau mapping between twist and phase}
At fixed pairing amplitudes, we write the phase-dependent GL energy as $F_\varphi=F_0-J_1\cos2\theta\cos\varphi+J_2\cos2\varphi$, where $J_1,J_2>0$ are the effective first- and second-harmonic Josephson energies~\cite{Can}. Their ratio $\mathcal K\equiv J_2/J_1$ sets the width of the TRSB window. Minimizing with respect to $\varphi$ gives equation~\eqref{eq:GL} for $|\cos2\theta|<4\mathcal K$; the critical angles satisfy $\cos2\theta_c^\mp =\pm4\mathcal K$. A fit to the BdG free energy gives $\mathcal{K}=0.136$ for $\Delta_1=40$~meV, $t=30$~meV (\Fig{fig:phasediagram}a).

\subsection*{Gaussian action of the collective modes}
We start from a fermionic action in the Nambu basis $\psi_{\mathbf{k}}$ with two pairing channels. The complex fields $\Delta_i(\tau)$, $i=1,2$, multiply the form factors $f_{\mathbf{k},i}$; at temperature $T$,
\begin{equation}
S = \sum_{i,j=1}^2 (U^{-1})_{ij}\int_0^{1/T}\!d\tau\,\Delta_i^*(\tau)\Delta_j(\tau)
-\sum_{\mathbf{k}}\int_0^{1/T}\!d\tau\,\psi_{\mathbf{k}}^\dagger(\tau)G^{-1}(\mathbf{k},\tau)\psi_{\mathbf{k}}(\tau),
\label{eq:action}
\end{equation}
where $U$ is the pairing-interaction matrix and $G^{-1}$ the inverse Gor'kov Green's function in the presence of the pairing fields. Integrating out the fermions gives $S_{\mathrm{eff}}=\sum_{ij}(U^{-1})_{ij}\int d\tau\,\Delta_i^*\Delta_j-\mathrm{Tr}\ln G^{-1}$, where $\mathrm{Tr}$ runs over Nambu indices, momentum and imaginary time. Expanding to quadratic order around the saddle point in the chiral basis
\begin{equation}
\chi_{\mathbf{k}}^{\pm} = (1 + r\cos 2\theta \cos\varphi)\,f_{d_{x^2-y^2}} \pm i\,r\sin 2\theta\,\sin\varphi\, f_{d_{xy}}
\label{eq:chi}
\end{equation}
gives, at zero momentum, the retarded fluctuation kernel in the $(h,a,b)$ basis,
\begin{equation}
\mathcal M^R(\Omega)=
\begin{pmatrix}
[D_h^R]^{-1}&-\Pi^{ha,R}/2&0\\
-\Pi^{ha,R}/2&[D_a^R]^{-1}&0\\
0&0&[D_b^R]^{-1}
\end{pmatrix},
\label{eq:kernel}
\end{equation}
with the diagonal inverse kernels and the Higgs--relative-amplitude mixing $\Pi^{ha,R}$ given in Supplementary Note~2. The relative-phase propagator is $D_b^R=1/\mathcal M_{bb}^R$; if $\lambda_\pm^R(\Omega)$ are the eigenvalues of the $(h,a)$ block, the hybrid amplitude propagators are $D_{A^\pm}^R=1/\lambda_\pm^R$. These full Gaussian propagators retain both the collective-mode poles and the two-quasiparticle continuum and are shown in \Fig{fig:modes}.

\subsection*{Mode energies across the TRSB window}
For the analysis of \Fig{fig:modes} we evaluate the kernels at $T=0$ and $\mathbf q=0$ in the continuum model with constant normal-state DOS $N_{\mathrm n}$. The integral over the band energy $\xi$ can be done analytically: with $x=\Omega/[2\Delta_0|\chi^+(\phi)|]$ and $F(x)=\arcsin x/(x\sqrt{1-x^2})$, each diagonal kernel reduces to
\begin{equation}
[D_i^R(\Omega)]^{-1}\propto\int_0^{2\pi}\frac{d\phi}{2\pi}\,|\chi^+|^2\,[w_i(\phi)-x^2]\,F(x),
\label{eq:Dangular}
\end{equation}
with $w_i=1$, $\cos^2 2\alpha$ and $\sin^2 2\alpha$ for $i=h,a,b$, and the mixing kernel to $\int\frac{d\phi}{2\pi}|\chi^+|^2\cos2\alpha\,(1-x^2)F(x)$. Roots below $2\Delta_{\min}$ are located by bracketing on a frequency grid refined logarithmically towards the threshold.This procedure reproduces the mode energies quoted in \Fig{fig:waterfall_int} and the mixing angle at $\varphi=\pi/2$. Where the binding energy falls below $10^{-3}$~meV we set $\Omega_i=2\Delta_{\min}$. The slopes in \Fig{fig:discriminant}c,d are linear fits of the resonance energies against $\Delta_{\min}$ over $\varphi\in[\pi/6,\pi/2]$.

\subsection*{Pole approximation and electron--boson coupling}
For the self-energy we retain only the isolated poles of $D_b$ and $D_{A^+}$, which lie below $2\Delta_{\min}$ in the parameter range considered, and neglect the non-resonant continuum. Expanding the inverse propagator at $\mathbf q=0$ to first order in $\nu^2-\Omega_i^2$ and assuming a factorized momentum profile,
\begin{equation}
    D_i^R(\nu, \mathbf{q}) \simeq \frac{Z_i}{2\Omega_i}\, F_i(\mathbf{q}) \left( \frac{1}{\nu - \Omega_i + i\delta} - \frac{1}{\nu + \Omega_i + i\delta} \right),
    \qquad F_i(\mathbf{q})=\frac{1}{1+\ell_i^2 \mathbf{q}^2},
    \label{eq:bosono_prop}
\end{equation}
where $i=b,A^+$, $Z_i^{-1}=\partial [D_i^R(\nu,\mathbf 0)]^{-1}/\partial(\nu^2)|_{\nu=\Omega_i}$ and $\ell_i(\varphi)=v_F/[\sqrt{2}\,\Omega_i(\varphi)]$ is the mode correlation length (Supplementary Note~3). This approximation is best justified when the pole is well separated from the continuum threshold; \Fig{fig:discriminant}b shows the binding energy across the TRSB window. Because $F_i$ is peaked at $\mathbf q=0$, the intermediate quasiparticle remains close to the same Fermi-surface angle, and the $\mathbf q$ integration sets the coupling strength
\begin{equation}
    U_i=\frac{Z_i}{2\Omega_i}\int\frac{d^2q}{(2\pi)^2}\,F_i(\mathbf q)\simeq\frac{Z_i\,\Omega_i}{4\pi v_F^2}.
    \label{eq:Ui}
\end{equation}

\subsection*{Self-energy and tunnelling density of states}
\begin{figure}[h]
  \centering
  \includegraphics[width=\textwidth]{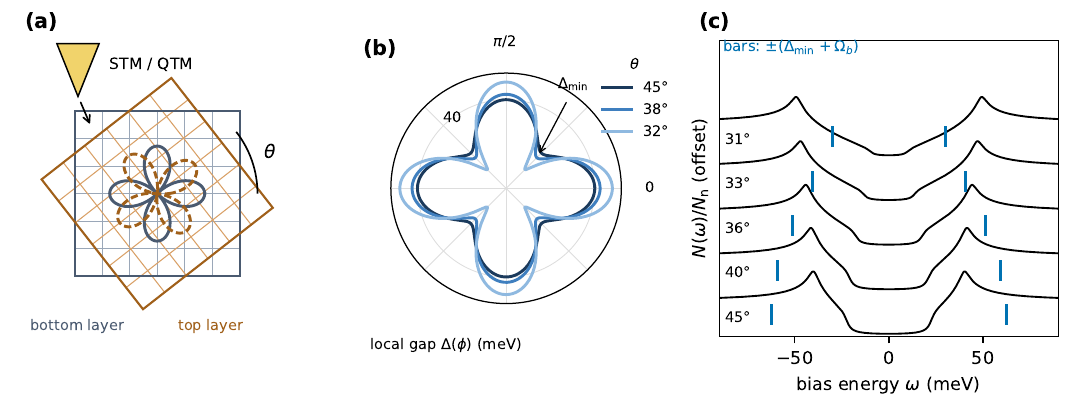}
  \caption{\textbf{Twist-tuned collective modes in tunnelling.}   \textbf{a}, Two Bi2212 layers twisted by $\theta$, with their $d$-wave form factors (solid, bottom layer; dashed, top layer), probed by a tunnelling tip (scanning tunnelling microscope, STM, or quantum twisting microscope, QTM). \textbf{b}, Local gap $\Delta(\phi)=\Delta_0|\chi^+(\phi)|$ on the Fermi surface in the chiral phase for three twist angles. The minimum $\Delta_{\min}$, at the top-layer nodal direction $\phi=\pi/4$, closes as $\theta$ moves away from $45^\circ$. \textbf{c}, Bare tunnelling DOS (no collective-mode self-energy) for several twist angles, offset vertically. Bars mark the predicted resonance energies $\pm(\Delta_{\min}+\Omega_b)$ of the relative-phase mode, which move with twist much faster than the coherence peaks. Parameters as in \Fig{fig:waterfall_int}; the twist angle is mapped onto the interlayer phase with equation~\eqref{eq:GL}.}
  \label{fig:concept}
\end{figure}

\begin{figure}[p]
  \centering
  \includegraphics[width=0.62\textwidth]{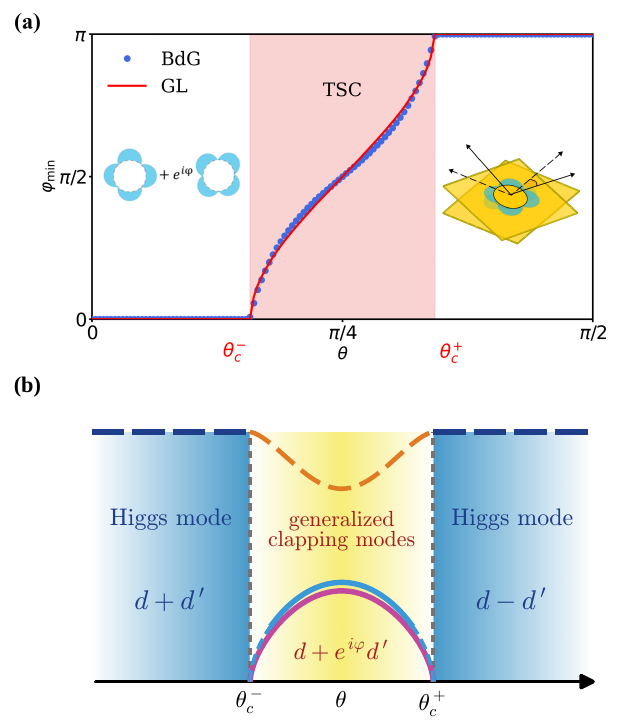}
  \caption{\textbf{Twisted double-layer $d$-wave superconductor.}   \textbf{a}, Relative phase between the layers, $\varphi_{\min}$, that minimizes the GL (line) and the BdG (symbols) free energy as a function of twist angle $\theta$, for unequal gaps in the two layers. Left inset: $\varphi_{\min}=\varphi$ is the relative phase angle between the two superconducting components. Right inset: Fermi surfaces and $d$-wave gaps in the continuum formulation of the microscopic model, and the twist angle $\theta$ between the layers.   \textbf{b}, Phase diagram. Below $\theta^-_c$ (above $\theta^+_c$) the system is in the $d\!+\!d'$ ($d\!-\!d'$) configuration; the ground state is then nodal and the only low-lying collective excitation is the Higgs mode. At the critical twist angles the system undergoes a superconducting-to-superconducting transition into the chiral phase, in which the bottom layer induces a minigap in the top layer and the collective modes become the clapping modes of a TRSB superconductor.   Parameters: $\Delta_1 = 40$~meV, $t=30$~meV, $\Delta_2 = t^2/\Delta_1$, $\mathcal{K} = 0.136$, $\theta_c^-=0.16\pi$, $\theta_c^+=0.34\pi$.}
  \label{fig:phasediagram}
\end{figure}

To leading order the self-energy arises from the exchange of a single collective mode,
\begin{equation}
\hat\Sigma(\mathbf{k},i\omega_n)
=T\sum_{i,m}\sum_{\mathbf q}D_i(\mathbf q,i\Omega_m)\,
\hat\Gamma_{i,\mathbf k}\,
\hat G_0(\mathbf{k}-\mathbf q,i\omega_n-i\Omega_m)\,
\hat\Gamma_{i,\mathbf{k}-\mathbf q},
    \label{eq:sigma}
\end{equation}
with fermionic $\omega_n$ and bosonic $\Omega_m$ Matsubara frequencies, bare Nambu Green's function $\hat G_0$, and vertices from equations~\eqref{eq:vertices} and \eqref{eq:GammaAplus}~\cite{PhysRevB.49.740,ZhuBalatsky2006}. We use a circular Fermi surface with $\Delta_{\mathbf k}\to\Delta_0\chi^+(\phi)$ and $\sum_{\mathbf k}\to N_{\mathrm n}\int d\xi \int_0^{2\pi}d\phi/2\pi$, and evaluate the vertices and the gap of $\hat G_0$ at the same angle $\phi$. After the Matsubara sum at $T=0$ and continuation $i\omega_n\to\omega+i0^+$, replaced by a small broadening $\gamma_\Sigma$ in the numerics, the retarded self-energy depends on momentum only through $\phi$ (Supplementary Note~3). The (1,1) component of the full Green's function is
\begin{equation}
G_{11}^R(\mathbf{k},\omega)
=\frac{z+\tilde\xi_{\mathbf{k}}-\Sigma_{22}^R}{(z-\tilde\xi_{\mathbf{k}}-\Sigma_{11}^R)(z+\tilde\xi_{\mathbf{k}}-\Sigma_{22}^R)-(\Delta_{\mathbf{k}}+\Sigma_{12}^R)(\Delta_{\mathbf{k}}^*+\Sigma_{21}^R)},
\label{eq:G11}
\end{equation}
with $z=\omega+i\Gamma_0$ and Dynes broadening $\Gamma_0$~\cite{PhysRevLett.41.1509}. The two anomalous factors are kept separately, since $\Sigma_{21}^R$ need not equal $(\Sigma_{12}^R)^*$ at real frequency. For the particle--hole-symmetric model ($\Sigma_{22}^R=\Sigma_{11}^R$), integration over $\xi$ gives
\begin{equation}
N(\omega)=-N_{\mathrm n}\!\int\!\frac{d\phi}{2\pi}\,
\operatorname{Im}\frac{\tilde\omega}{\sqrt{\tilde\Delta^{+}\tilde\Delta^{-}-\tilde\omega^{2}}},\qquad
\tilde\omega=\omega+i\Gamma_0-\Sigma_{11}^R,\quad
\tilde\Delta^\pm=\Delta_0\chi^\pm+\Sigma_{12/21}^R,
\label{eq:Ak}
\end{equation}
with the square-root branch chosen so that the DOS is positive and approaches $N_{\mathrm n}$ at large $|\omega|$. The integrand, multiplied by $-N_{\mathrm n}$, defines the angle-resolved DOS $N(\phi,\omega)$, whose angular average is the tunnelling DOS. Without self-energy, equation~\eqref{eq:Ak} gives $N_{\mathrm{bare}}(\omega)$, which is used in \Fig{fig:concept}c.

\subsection*{Use of AI tools}
Claude (Anthropic) was used to check analytical calculations, to assist in developing the code used for the numerical analysis and for Figs.~1, 3 and 7 and Supplementary Fig.~3, and to improve the language of the manuscript. The authors reviewed and verified all AI-assisted output and take full responsibility for the content of the paper.



\bibliographystyle{naturemagdoi}
\bibliography{refs}

\begin{thebibliography}{10}
\expandafter\ifx\csname url\endcsname\relax
  \def\url#1{\texttt{#1}}\fi
\expandafter\ifx\csname urlprefix\endcsname\relax\def\urlprefix{Available at
  }\fi
\providecommand{\bibinfo}[2]{#2}
\providecommand{\eprint}[2][]{\url{#2}}
\makeatletter
\providecommand \doibase [0]{https://doi.org/}
\providecommand \@sanitize@url [0]{\catcode `\\12\catcode `\$12\catcode
  `\&12\catcode `\#12\catcode `\^12\catcode `\_12\catcode `\%12\relax}
\providecommand \Doi [0]{\begingroup \@sanitize@url \@Doi }%
\providecommand \@Doi [1]{\endgroup\@@startlink{\doibase#1}\@@Doi}%
\def\gobbledotblock , {}
\let\@@Doi\relax
\newcommand \@@Doi
  [1]{\textcolor{docnotelinkcolor}{\gobbledotblock#1}\@@endlink}
\newcommand\myhrefnoop[2]{\gobbledotblock#2}
\newcommand\@@myhref[2]{\href{#1}{\gobbledotblock#2}}
\newcommand\@myhref[1]{\endgroup\@@myhref{#1}}
\newcommand\myhref{\begingroup\@sanitize@url\@myhref}
\providecommand\@@startlink[1]{%
\leavevmode%
\pdfstartlink\pdfstartlink@attr%
 user{/Subtype/Link/A<</Type/Action/S/URI/URI(#1)>>}\relax%
}
\providecommand\@@endlink{\pdfendlink}
\providecommand\pdfstartlink@attr{attr{/Border[0 0 0 [1 5] ]/H/I/C[0 1 1]}}
\makeatother


\bibitem{PekkerVarma2015}
\bibinfo{author}{Pekker, D.} \& \bibinfo{author}{Varma, C.~M.}
\newblock \bibinfo{title}{{Amplitude/Higgs Modes in Condensed Matter Physics}}.
\newblock \Doi {10.1146/annurev-conmatphys-031214-014350} {,
  \emph{\bibinfo{journal}{Annual Review of Condensed Matter Physics}}
  \textbf{\bibinfo{volume}{6}}, \bibinfo{pages}{269--297}
  (\bibinfo{year}{2015})}.

\bibitem{ShimanoTsuji2020}
\bibinfo{author}{Shimano, R.} \& \bibinfo{author}{Tsuji, N.}
\newblock \bibinfo{title}{{Higgs Mode in Superconductors}}.
\newblock \Doi {10.1146/annurev-conmatphys-031119-050813} {,
  \emph{\bibinfo{journal}{Annual Review of Condensed Matter Physics}}
  \textbf{\bibinfo{volume}{11}}, \bibinfo{pages}{103--124}
  (\bibinfo{year}{2020})}.

\bibitem{BarlasVarma2013}
\bibinfo{author}{Barlas, Y.} \& \bibinfo{author}{Varma, C.~M.}
\newblock \bibinfo{title}{{Amplitude or Higgs modes in d-wave
  superconductors}}.
\newblock \Doi {10.1103/PhysRevB.87.054503} {, \emph{\bibinfo{journal}{Phys.
  Rev. B}} \textbf{\bibinfo{volume}{87}}, \bibinfo{pages}{054503}
  (\bibinfo{year}{2013})}.

\bibitem{Matsunaga2013}
\bibinfo{author}{Matsunaga, R.} \emph{et~al.}
\newblock \bibinfo{title}{{Higgs Amplitude Mode in the BCS Superconductors
  {Nb$_{1-x}$Ti$_x$N} Induced by Terahertz Pulse Excitation}}.
\newblock \Doi {10.1103/PhysRevLett.111.057002} {,
  \emph{\bibinfo{journal}{Phys. Rev. Lett.}} \textbf{\bibinfo{volume}{111}},
  \bibinfo{pages}{057002} (\bibinfo{year}{2013})}.

\bibitem{Matsunaga2014}
\bibinfo{author}{Matsunaga, R.} \emph{et~al.}
\newblock \bibinfo{title}{{Light-induced collective pseudospin precession
  resonating with Higgs mode in a superconductor}}.
\newblock \Doi {10.1126/science.1254697} {, \emph{\bibinfo{journal}{Science}}
  \textbf{\bibinfo{volume}{345}}, \bibinfo{pages}{1145--1149}
  (\bibinfo{year}{2014})}.

\bibitem{Measson2014}
\bibinfo{author}{M\'easson, M.-A.} \emph{et~al.}
\newblock \bibinfo{title}{{Amplitude Higgs mode in the {2H-NbSe$_2$}
  superconductor}}.
\newblock \Doi {10.1103/PhysRevB.89.060503} {, \emph{\bibinfo{journal}{Phys.
  Rev. B}} \textbf{\bibinfo{volume}{89}}, \bibinfo{pages}{060503}
  (\bibinfo{year}{2014})}.

\bibitem{PhysRevLett.120.117001}
\bibinfo{author}{Katsumi, K.} \emph{et~al.}
\newblock \bibinfo{title}{{Higgs Mode in the $d$-Wave Superconductor
  ${\mathrm{Bi}}_{2}{\mathrm{Sr}}_{2}{\mathrm{CaCu}}_{2}{\mathrm{O}}_{8+x}$
  Driven by an Intense Terahertz Pulse}}.
\newblock \Doi {10.1103/PhysRevLett.120.117001} {,
  \emph{\bibinfo{journal}{Phys. Rev. Lett.}} \textbf{\bibinfo{volume}{120}},
  \bibinfo{pages}{117001} (\bibinfo{year}{2018})}.

\bibitem{Leggett1966}
\bibinfo{author}{Leggett, A.~J.}
\newblock \bibinfo{title}{{Number-Phase Fluctuations in Two-Band
  Superconductors}}.
\newblock \Doi {10.1143/PTP.36.901} {, \emph{\bibinfo{journal}{Progress of
  Theoretical Physics}} \textbf{\bibinfo{volume}{36}},
  \bibinfo{pages}{901--930} (\bibinfo{year}{1966})}.

\bibitem{PhysRevLett.99.227002}
\bibinfo{author}{Blumberg, G.} \emph{et~al.}
\newblock \bibinfo{title}{{Observation of Leggett's Collective Mode in a
  Multiband ${\mathrm{MgB}}_{2}$ Superconductor}}.
\newblock \Doi {10.1103/PhysRevLett.99.227002} {, \emph{\bibinfo{journal}{Phys.
  Rev. Lett.}} \textbf{\bibinfo{volume}{99}}, \bibinfo{pages}{227002}
  (\bibinfo{year}{2007})}.

\bibitem{BardasisSchrieffer1961}
\bibinfo{author}{Bardasis, A.} \& \bibinfo{author}{Schrieffer, J.~R.}
\newblock \bibinfo{title}{{Excitons and Plasmons in Superconductors}}.
\newblock \Doi {10.1103/PhysRev.121.1050} {, \emph{\bibinfo{journal}{Phys.
  Rev.}} \textbf{\bibinfo{volume}{121}}, \bibinfo{pages}{1050--1062}
  (\bibinfo{year}{1961})}.

\bibitem{PhysRevB.92.094506}
\bibinfo{author}{Maiti, S.} \& \bibinfo{author}{Hirschfeld, P.~J.}
\newblock \bibinfo{title}{{Collective modes in superconductors with competing
  $s$- and $d$-wave interactions}}.
\newblock \Doi {10.1103/PhysRevB.92.094506} {, \emph{\bibinfo{journal}{Phys.
  Rev. B}} \textbf{\bibinfo{volume}{92}}, \bibinfo{pages}{094506}
  (\bibinfo{year}{2015})}.

\bibitem{PhysRevX.4.041046}
\bibinfo{author}{B{\"o}hm, T.} \emph{et~al.}
\newblock \bibinfo{title}{{Balancing Act: Evidence for a Strong Subdominant
  $d$-Wave Pairing Channel in
  ${\mathrm{Ba}}_{0.6}{\mathrm{K}}_{0.4}{\mathrm{Fe}}_{2}{\mathrm{As}}_{2}$}}.
\newblock \Doi {10.1103/PhysRevX.4.041046} {, \emph{\bibinfo{journal}{Phys.
  Rev. X}} \textbf{\bibinfo{volume}{4}}, \bibinfo{pages}{041046}
  (\bibinfo{year}{2014})}.

\bibitem{Balatsky_2000}
\bibinfo{author}{Balatsky, A.}, \bibinfo{author}{Kumar, P.} \&
  \bibinfo{author}{Schrieffer, J.}
\newblock \bibinfo{title}{{Clapping modes in unconventional superconductors}}.
\newblock \Doi {10.1016/s0921-4534(00)00700-0} {,
  \emph{\bibinfo{journal}{Physica C: Superconductivity}}
  \textbf{\bibinfo{volume}{341-348}}, \bibinfo{pages}{807--810}
  (\bibinfo{year}{2000})}.

\bibitem{SaulsWuChung2015}
\bibinfo{author}{Sauls, J.~A.}, \bibinfo{author}{Wu, H.} \&
  \bibinfo{author}{Chung, S.~B.}
\newblock \bibinfo{title}{{Anisotropy and Strong-Coupling Effects on the
  Collective Mode Spectrum of Chiral Superconductors: Application to
  {Sr$_2$RuO$_4$}}}.
\newblock \Doi {10.3389/fphy.2015.00036} {, \emph{\bibinfo{journal}{Frontiers
  in Physics}} \textbf{\bibinfo{volume}{3}}, \bibinfo{pages}{36}
  (\bibinfo{year}{2015})}.

\bibitem{Poniatowski2022}
\bibinfo{author}{Poniatowski, N.~R.}, \bibinfo{author}{Curtis, J.~B.},
  \bibinfo{author}{Yacoby, A.} \& \bibinfo{author}{Narang, P.}
\newblock \bibinfo{title}{{Spectroscopic signatures of time-reversal symmetry
  breaking superconductivity}}.
\newblock \Doi {10.1038/s42005-022-00819-0} {,
  \emph{\bibinfo{journal}{Communications Physics}}
  \textbf{\bibinfo{volume}{5}}, \bibinfo{pages}{44} (\bibinfo{year}{2022})}.

\bibitem{Bardeen1961}
\bibinfo{author}{Bardeen, J.}
\newblock \bibinfo{title}{{Tunnelling from a Many-Particle Point of View}}.
\newblock \Doi {10.1103/PhysRevLett.6.57} {, \emph{\bibinfo{journal}{Phys. Rev.
  Lett.}} \textbf{\bibinfo{volume}{6}}, \bibinfo{pages}{57--59}
  (\bibinfo{year}{1961})}.

\bibitem{TersoffHamann1983}
\bibinfo{author}{Tersoff, J.} \& \bibinfo{author}{Hamann, D.~R.}
\newblock \bibinfo{title}{{Theory and Application for the Scanning Tunneling
  Microscope}}.
\newblock \Doi {10.1103/PhysRevLett.50.1998} {, \emph{\bibinfo{journal}{Phys.
  Rev. Lett.}} \textbf{\bibinfo{volume}{50}}, \bibinfo{pages}{1998--2001}
  (\bibinfo{year}{1983})}.

\bibitem{Fischer2007}
\bibinfo{author}{Fischer, {\O}.}, \bibinfo{author}{Kugler, M.},
  \bibinfo{author}{Maggio-Aprile, I.}, \bibinfo{author}{Berthod, C.} \&
  \bibinfo{author}{Renner, C.}
\newblock \bibinfo{title}{{Scanning tunneling spectroscopy of high-temperature
  superconductors}}.
\newblock \Doi {10.1103/RevModPhys.79.353} {, \emph{\bibinfo{journal}{Rev. Mod.
  Phys.}} \textbf{\bibinfo{volume}{79}}, \bibinfo{pages}{353--419}
  (\bibinfo{year}{2007})}.

\bibitem{Eliashberg1960}
\bibinfo{author}{Eliashberg, G.~M.}
\newblock \bibinfo{title}{{Interactions between electrons and lattice
  vibrations in a superconductor}}.
\newblock \myhrefnoop {} {, \emph{\bibinfo{journal}{Sov. Phys. JETP}}
  \textbf{\bibinfo{volume}{11}}, \bibinfo{pages}{696--702}
  (\bibinfo{year}{1960})}.

\bibitem{ScalapinoSchriefferWilkins1966}
\bibinfo{author}{Scalapino, D.~J.}, \bibinfo{author}{Schrieffer, J.~R.} \&
  \bibinfo{author}{Wilkins, J.~W.}
\newblock \bibinfo{title}{{Strong-Coupling Superconductivity. I}}.
\newblock \Doi {10.1103/PhysRev.148.263} {, \emph{\bibinfo{journal}{Phys.
  Rev.}} \textbf{\bibinfo{volume}{148}}, \bibinfo{pages}{263--279}
  (\bibinfo{year}{1966})}.

\bibitem{McMillanRowell1965}
\bibinfo{author}{McMillan, W.~L.} \& \bibinfo{author}{Rowell, J.~M.}
\newblock \bibinfo{title}{{Lead Phonon Spectrum Calculated from Superconducting
  Density of States}}.
\newblock \Doi {10.1103/PhysRevLett.14.108} {, \emph{\bibinfo{journal}{Phys.
  Rev. Lett.}} \textbf{\bibinfo{volume}{14}}, \bibinfo{pages}{108--112}
  (\bibinfo{year}{1965})}.

\bibitem{Chubukov2003}
\bibinfo{author}{Chubukov, A.~V.}, \bibinfo{author}{Pines, D.} \&
  \bibinfo{author}{Schmalian, J.}
\newblock \emph{\bibinfo{title}{{A Spin Fluctuation Model for d-Wave
  Superconductivity}}}, \bibinfo{pages}{495--590} (\bibinfo{publisher}{Springer
  Berlin Heidelberg}, \bibinfo{address}{Berlin, Heidelberg},
  \bibinfo{year}{2003}).

\bibitem{PhysRevLett.87.067005}
\bibinfo{author}{Zasadzinski, J.~F.} \emph{et~al.}
\newblock \bibinfo{title}{{Correlation of Tunneling Spectra in
  ${\mathrm{Bi}}_{2}{\mathrm{Sr}}_{2}{\mathrm{CaCu}}_{2}{O}_{8+\ensuremath{\delta}}$
  with the Resonance Spin Excitation}}.
\newblock \Doi {10.1103/PhysRevLett.87.067005} {, \emph{\bibinfo{journal}{Phys.
  Rev. Lett.}} \textbf{\bibinfo{volume}{87}}, \bibinfo{pages}{067005}
  (\bibinfo{year}{2001})}.

\bibitem{Zasadzinski2006}
\bibinfo{author}{Zasadzinski, J.~F.}, \bibinfo{author}{Coffey, L.},
  \bibinfo{author}{Romano, P.} \& \bibinfo{author}{Yusof, Z.}
\newblock \bibinfo{title}{{Persistence of Strong Electron Coupling to a Narrow
  Boson Spectrum in Overdoped Bi$_2$Sr$_2$CaCu$_2$O$_{8+\delta}$ Tunneling
  Data}}.
\newblock \Doi {10.1103/PhysRevLett.96.017004} {, \emph{\bibinfo{journal}{Phys.
  Rev. Lett.}} \textbf{\bibinfo{volume}{96}}, \bibinfo{pages}{017004}
  (\bibinfo{year}{2006})}.

\bibitem{PhysRevLett.85.3261}
\bibinfo{author}{Eschrig, M.} \& \bibinfo{author}{Norman, M.~R.}
\newblock \bibinfo{title}{{Neutron Resonance: Modeling Photoemission and
  Tunneling Data in the Superconducting State of
  ${{\mathrm{Bi}}_{2}{\mathrm{Sr}}_{2}{\mathrm{CaCu}}_{2}O}_{8+\mathit{\ensuremath{\delta}}}$}}.
\newblock \Doi {10.1103/PhysRevLett.85.3261} {, \emph{\bibinfo{journal}{Phys.
  Rev. Lett.}} \textbf{\bibinfo{volume}{85}}, \bibinfo{pages}{3261--3264}
  (\bibinfo{year}{2000})}.

\bibitem{PhysRevLett.83.1652}
\bibinfo{author}{Abanov, A.} \& \bibinfo{author}{Chubukov, A.~V.}
\newblock \bibinfo{title}{{A Relation between the Resonance Neutron Peak and
  ARPES Data in Cuprates}}.
\newblock \Doi {10.1103/PhysRevLett.83.1652} {, \emph{\bibinfo{journal}{Phys.
  Rev. Lett.}} \textbf{\bibinfo{volume}{83}}, \bibinfo{pages}{1652--1655}
  (\bibinfo{year}{1999})}.

\bibitem{Devereaux2004}
\bibinfo{author}{Devereaux, T.~P.}, \bibinfo{author}{Cuk, T.},
  \bibinfo{author}{Shen, Z.-X.} \& \bibinfo{author}{Nagaosa, N.}
\newblock \bibinfo{title}{{Anisotropic Electron-Phonon Interaction in the
  Cuprates}}.
\newblock \Doi {10.1103/PhysRevLett.93.117004} {, \emph{\bibinfo{journal}{Phys.
  Rev. Lett.}} \textbf{\bibinfo{volume}{93}}, \bibinfo{pages}{117004}
  (\bibinfo{year}{2004})}.

\bibitem{Shan2012}
\bibinfo{author}{Shan, L.} \emph{et~al.}
\newblock \bibinfo{title}{{Evidence of a Spin Resonance Mode in the Iron-Based
  Superconductor Ba$_{0.6}$K$_{0.4}$Fe$_2$As$_2$ from Scanning Tunneling
  Spectroscopy}}.
\newblock \Doi {10.1103/PhysRevLett.108.227002} {,
  \emph{\bibinfo{journal}{Phys. Rev. Lett.}} \textbf{\bibinfo{volume}{108}},
  \bibinfo{pages}{227002} (\bibinfo{year}{2012})}.

\bibitem{Levy2007}
\bibinfo{author}{Levy~de Castro, G.}, \bibinfo{author}{Berthod, C.},
  \bibinfo{author}{Piriou, A.}, \bibinfo{author}{Giannini, E.} \&
  \bibinfo{author}{Fischer, O.}
\newblock \bibinfo{title}{{Preeminent Role of the Van Hove Singularity in the
  Strong-Coupling Analysis of Scanning Tunneling Spectroscopy for
  Two-Dimensional Cuprate Superconductors}}.
\newblock \Doi {10.1103/PhysRevLett.101.267004} {,
  \emph{\bibinfo{journal}{Phys. Rev. Lett.}} \textbf{\bibinfo{volume}{101}},
  \bibinfo{pages}{267004} (\bibinfo{year}{2008})}.

\bibitem{Lee2006}
\bibinfo{author}{Lee, J.} \emph{et~al.}
\newblock \bibinfo{title}{{Interplay of electron-lattice interactions and
  superconductivity in {Bi$_2$Sr$_2$CaCu$_2$O$_{8+\delta}$}}}.
\newblock \Doi {10.1038/nature04973} {, \emph{\bibinfo{journal}{Nature}}
  \textbf{\bibinfo{volume}{442}}, \bibinfo{pages}{546--550}
  (\bibinfo{year}{2006})}.

\bibitem{Ferrell1969}
\bibinfo{author}{Ferrell, R.~A.}
\newblock \bibinfo{title}{Fluctuations and the superconducting phase
  transition. ii. onset of josephson tunneling and paraconductivity of a
  junction}.
\newblock \Doi {10.1007/BF00628207} {, \emph{\bibinfo{journal}{J. Low Temp.
  Phys.; (United States)}} \textbf{\bibinfo{volume}{1}}
  (\bibinfo{year}{1969})}.
\newblock \urlprefix\url{https://www.osti.gov/biblio/7184105}.

\bibitem{PhysRevLett.24.1052}
\bibinfo{author}{Scalapino, D.~J.}
\newblock \bibinfo{title}{{Pair Tunneling as a Probe of Fluctuations in
  Superconductors}}.
\newblock \Doi {10.1103/PhysRevLett.24.1052} {, \emph{\bibinfo{journal}{Phys.
  Rev. Lett.}} \textbf{\bibinfo{volume}{24}}, \bibinfo{pages}{1052--1055}
  (\bibinfo{year}{1970})}.

\bibitem{PhysRevB.49.740}
\bibinfo{author}{Gherghetta, T.} \& \bibinfo{author}{Nambu, Y.}
\newblock \bibinfo{title}{{Feedback effects in superconductors}}.
\newblock \Doi {10.1103/PhysRevB.49.740} {, \emph{\bibinfo{journal}{Phys. Rev.
  B}} \textbf{\bibinfo{volume}{49}}, \bibinfo{pages}{740(R)--743(R)}
  (\bibinfo{year}{1994})}.

\bibitem{LeeSteiner2023}
\bibinfo{author}{Lee, P.~A.} \& \bibinfo{author}{Steiner, J.~F.}
\newblock \bibinfo{title}{{Detection of collective modes in unconventional
  superconductors using tunneling spectroscopy}}.
\newblock \Doi {10.1103/PhysRevB.108.174503} {, \emph{\bibinfo{journal}{Phys.
  Rev. B}} \textbf{\bibinfo{volume}{108}}, \bibinfo{pages}{174503}
  (\bibinfo{year}{2023})}.

\bibitem{Wan2022NbSe2}
\bibinfo{author}{Wan, W.} \emph{et~al.}
\newblock \bibinfo{title}{{Observation of Superconducting Collective Modes from
  Competing Pairing Instabilities in Single-Layer NbSe2}}.
\newblock \Doi {10.1002/adma.202206078} {, \emph{\bibinfo{journal}{Advanced
  Materials}} \textbf{\bibinfo{volume}{34}}, \bibinfo{pages}{2206078}
  (\bibinfo{year}{2022})}.

\bibitem{Hu2024}
\bibinfo{author}{Hu, B.} \emph{et~al.}
\newblock \bibinfo{title}{{Evidence of a distinct collective mode in Kagome
  superconductors}}.
\newblock \Doi {10.1038/s41467-024-50330-z} {, \emph{\bibinfo{journal}{Nature
  Communications}} \textbf{\bibinfo{volume}{15}}, \bibinfo{pages}{6109}
  (\bibinfo{year}{2024})}.

\bibitem{Can}
\bibinfo{author}{Can, O.} \emph{et~al.}
\newblock \bibinfo{title}{{High-temperature topological superconductivity in
  twisted double-layer copper oxides}}.
\newblock \Doi {10.1038/s41567-020-01142-7} {, \emph{\bibinfo{journal}{Nature
  Physics}} \textbf{\bibinfo{volume}{17}}, \bibinfo{pages}{519--524}
  (\bibinfo{year}{2021})}.

\bibitem{PhysRevB.105.064501}
\bibinfo{author}{Tummuru, T.}, \bibinfo{author}{Plugge, S.} \&
  \bibinfo{author}{Franz, M.}
\newblock \bibinfo{title}{{Josephson effects in twisted cuprate bilayers}}.
\newblock \Doi {10.1103/PhysRevB.105.064501} {, \emph{\bibinfo{journal}{Phys.
  Rev. B}} \textbf{\bibinfo{volume}{105}}, \bibinfo{pages}{064501}
  (\bibinfo{year}{2022})}.

\bibitem{PhysRevB.109.045111}
\bibinfo{author}{B\'elanger, M.} \& \bibinfo{author}{S\'en\'echal, D.}
\newblock \bibinfo{title}{{Doping dependence of chiral superconductivity in
  near ${45}^{\ensuremath{\circ}}$ twisted bilayer cuprates}}.
\newblock \Doi {10.1103/PhysRevB.109.045111} {, \emph{\bibinfo{journal}{Phys.
  Rev. B}} \textbf{\bibinfo{volume}{109}}, \bibinfo{pages}{045111}
  (\bibinfo{year}{2024})}.

\bibitem{Song2022}
\bibinfo{author}{Song, X.-Y.}, \bibinfo{author}{Zhang, Y.-H.} \&
  \bibinfo{author}{Vishwanath, A.}
\newblock \bibinfo{title}{{Doping a moir\'e Mott insulator: A $t$-$J$ model
  study of twisted cuprates}}.
\newblock \Doi {10.1103/PhysRevB.105.L201102} {, \emph{\bibinfo{journal}{Phys.
  Rev. B}} \textbf{\bibinfo{volume}{105}}, \bibinfo{pages}{L201102}
  (\bibinfo{year}{2022})}.

\bibitem{Volkov2025}
\bibinfo{author}{Volkov, P.~A.} \emph{et~al.}
\newblock \bibinfo{title}{{Josephson effects in twisted nodal
  superconductors}}.
\newblock \Doi {10.1103/PhysRevB.111.014514} {, \emph{\bibinfo{journal}{Phys.
  Rev. B}} \textbf{\bibinfo{volume}{111}}, \bibinfo{pages}{014514}
  (\bibinfo{year}{2025})}.

\bibitem{Zhao2023}
\bibinfo{author}{Zhao, S. Y.~F.} \emph{et~al.}
\newblock \bibinfo{title}{{Time-reversal symmetry breaking superconductivity
  between twisted cuprate superconductors}}.
\newblock \Doi {10.1126/science.abl8371} {, \emph{\bibinfo{journal}{Science}}
  \textbf{\bibinfo{volume}{382}}, \bibinfo{pages}{1422--1427}
  (\bibinfo{year}{2023})}.

\bibitem{Zhu2021}
\bibinfo{author}{Zhu, Y.} \emph{et~al.}
\newblock \bibinfo{title}{{Presence of $s$-wave pairing in {J}osephson
  junctions made of twisted ultrathin {Bi$_2$Sr$_2$CaCu$_2$O$_{8+x}$} flakes}}.
\newblock \Doi {10.1103/PhysRevX.11.031011} {, \emph{\bibinfo{journal}{Phys.
  Rev. X}} \textbf{\bibinfo{volume}{11}}, \bibinfo{pages}{031011}
  (\bibinfo{year}{2021})}.

\bibitem{Martini2023}
\bibinfo{author}{Martini, M.} \emph{et~al.}
\newblock \bibinfo{title}{{Twisted cuprate van der {W}aals heterostructures
  with controlled {J}osephson coupling}}.
\newblock \myhrefnoop {} {, \emph{\bibinfo{journal}{Mater. Today}}
  \textbf{\bibinfo{volume}{67}} (\bibinfo{year}{2023})}.

\bibitem{PhysRevLett.132.017002}
\bibinfo{author}{Patel, H.}, \bibinfo{author}{Pathak, V.},
  \bibinfo{author}{Can, O.}, \bibinfo{author}{Potter, A.~C.} \&
  \bibinfo{author}{Franz, M.}
\newblock \bibinfo{title}{{$d$-Mon: A Transmon with Strong Anharmonicity Based
  on Planar $c$-Axis Tunneling Junction between $d$-Wave and $s$-Wave
  Superconductors}}.
\newblock \Doi {10.1103/PhysRevLett.132.017002} {,
  \emph{\bibinfo{journal}{Phys. Rev. Lett.}} \textbf{\bibinfo{volume}{132}},
  \bibinfo{pages}{017002} (\bibinfo{year}{2024})}.

\bibitem{PhysRevB.53.3598}
\bibinfo{author}{Musaelian, K.~A.}, \bibinfo{author}{Betouras, J.},
  \bibinfo{author}{Chubukov, A.~V.} \& \bibinfo{author}{Joynt, R.}
\newblock \bibinfo{title}{{Mixed-symmetry superconductivity in two-dimensional
  Fermi liquids}}.
\newblock \Doi {10.1103/PhysRevB.53.3598} {, \emph{\bibinfo{journal}{Phys. Rev.
  B}} \textbf{\bibinfo{volume}{53}}, \bibinfo{pages}{3598--3603}
  (\bibinfo{year}{1996})}.

\bibitem{Inbar2023}
\bibinfo{author}{Inbar, A.} \emph{et~al.}
\newblock \bibinfo{title}{{The quantum twisting microscope}}.
\newblock \Doi {10.1038/s41586-022-05685-y} {, \emph{\bibinfo{journal}{Nature}}
  \textbf{\bibinfo{volume}{614}}, \bibinfo{pages}{682--687}
  (\bibinfo{year}{2023})}.

\bibitem{Pan2001}
\bibinfo{author}{Pan, S.~H.} \emph{et~al.}
\newblock \bibinfo{title}{{Microscopic electronic inhomogeneity in the
  high-{$T_c$} superconductor {Bi$_2$Sr$_2$CaCu$_2$O$_{8+x}$}}}.
\newblock \Doi {10.1038/35095012} {, \emph{\bibinfo{journal}{Nature}}
  \textbf{\bibinfo{volume}{413}}, \bibinfo{pages}{282--285}
  (\bibinfo{year}{2001})}.

\bibitem{ZhuBalatsky2006}
\bibinfo{author}{Zhu, J.-X.} \emph{et~al.}
\newblock \bibinfo{title}{{Fourier-transformed local density of states and
  tunneling into a d-wave superconductor with bosonic modes}}.
\newblock \Doi {10.1103/PhysRevB.73.014511} {, \emph{\bibinfo{journal}{Phys.
  Rev. B}} \textbf{\bibinfo{volume}{73}}, \bibinfo{pages}{014511}
  (\bibinfo{year}{2006})}.

\bibitem{PhysRevLett.41.1509}
\bibinfo{author}{Dynes, R.~C.}, \bibinfo{author}{Narayanamurti, V.} \&
  \bibinfo{author}{Garno, J.~P.}
\newblock \bibinfo{title}{{Direct Measurement of Quasiparticle-Lifetime
  Broadening in a Strong-Coupled Superconductor}}.
\newblock \Doi {10.1103/PhysRevLett.41.1509} {, \emph{\bibinfo{journal}{Phys.
  Rev. Lett.}} \textbf{\bibinfo{volume}{41}}, \bibinfo{pages}{1509--1512}
  (\bibinfo{year}{1978})}.

\bibitem{JURECKA1999304}
\bibinfo{author}{Jurecka, C.} \& \bibinfo{author}{Schachinger, E.}
\newblock \bibinfo{title}{{The phase diagram of an orthorhombic d-wave
  superconductor}}.
\newblock \Doi {https://doi.org/10.1016/S0921-4534(98)00694-7} {,
  \emph{\bibinfo{journal}{Physica C: Superconductivity}}
  \textbf{\bibinfo{volume}{312}}, \bibinfo{pages}{304--312}
  (\bibinfo{year}{1999})}.

\bibitem{thz_near}
\bibinfo{author}{Sun, Z.}, \bibinfo{author}{Fogler, M.~M.},
  \bibinfo{author}{Basov, D.~N.} \& \bibinfo{author}{Millis, A.~J.}
\newblock \bibinfo{title}{{Collective modes and terahertz near-field response
  of superconductors}}.
\newblock \Doi {10.1103/PhysRevResearch.2.023413} {,
  \emph{\bibinfo{journal}{Phys. Rev. Res.}} \textbf{\bibinfo{volume}{2}},
  \bibinfo{pages}{023413} (\bibinfo{year}{2020})}.

\end{thebibliography}

\section*{Acknowledgements}
We thank the Max Planck--UBC--UTokyo Center for Quantum Materials for fruitful collaborations and financial support. Work at UBC was supported by NSERC and CIFAR.

\section*{Author contributions}
S.N. performed the calculations. M.F. and D.V. supervised the project. All authors discussed the results and contributed to writing the manuscript.

\section*{Competing interests}
The authors declare no competing interests.

\clearpage
\section*{Figures}
\vspace{100pt}

\begin{figure}[p]
  \centering
  \includegraphics[width=\textwidth]{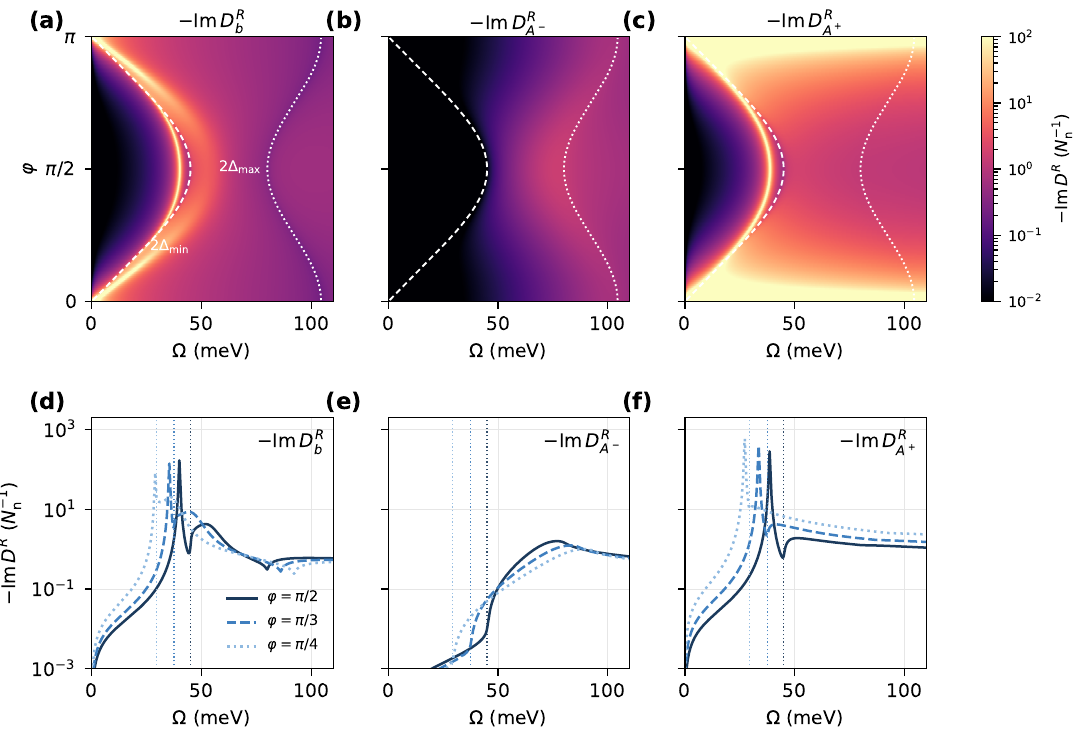}
  \caption{\textbf{Collective modes of the chiral order parameter $\Delta_{\mathbf{k}}=\Delta_0\chi^+_{\mathbf{k}}$.}
  \textbf{a}--\textbf{c}, Full Gaussian spectral functions $-\mathrm{Im}\,D_i^R(\Omega,\varphi)$ (logarithmic colour scale) as functions of frequency $\Omega$ and equilibrium relative phase $\varphi$ for (\textbf{a}) the relative-phase mode $D_b$, (\textbf{b}) the amplitude mode $D_{A^-}$ and (\textbf{c}) the amplitude mode $D_{A^+}$, computed from the frequency-dependent fluctuation kernels of Supplementary Note~2. The dashed curve is the two-particle threshold $2\Delta_{\min}(\varphi)$ and the dotted curve marks $2\Delta_{\max}(\varphi)$. A mode below the dashed curve is undamped, whereas a mode within the continuum can decay into quasiparticle pairs.
  \textbf{d}--\textbf{f}, Frequency cuts of the corresponding spectral functions at $\varphi=\pi/2$, $\pi/3$ and $\pi/4$; vertical dotted lines mark $2\Delta_{\min}$. The relative-phase mode (\textbf{d}) and the $A^+$ amplitude mode (\textbf{f}) are sharp resonances below $2\Delta_{\min}$, at $\Omega_b=39.9$~meV and $\Omega_{A^+}=38.5$~meV for $\varphi=\pi/2$, whereas the $A^-$ amplitude mode (\textbf{e}) remains broad and damped.
  The amplitude spectral functions are projections of the $2\times2$ amplitude propagator onto the normal-mode vector $(\cos\beta_{A^+},\sin\beta_{A^+})$ at the $A^+$ pole and onto its orthogonal complement. Close to $\varphi=0$ and $\pi$, where the two chiral components coincide and the chiral order vanishes, the $A^+$ channel softens and its spectral weight grows strongly (saturated region in \textbf{c}).
  Parameters: $\Delta_1=40$~meV, $t=30$~meV, $\Delta_2=t^2/\Delta_1$, $\mathcal{K}=0.136$; retarded broadening $\eta=0.25$~meV. Spectral functions are in units of $N_{\mathrm n}^{-1}$.}
  \label{fig:modes}
\end{figure}

\begin{figure}[p]
  \centering
  \includegraphics[width=\textwidth]{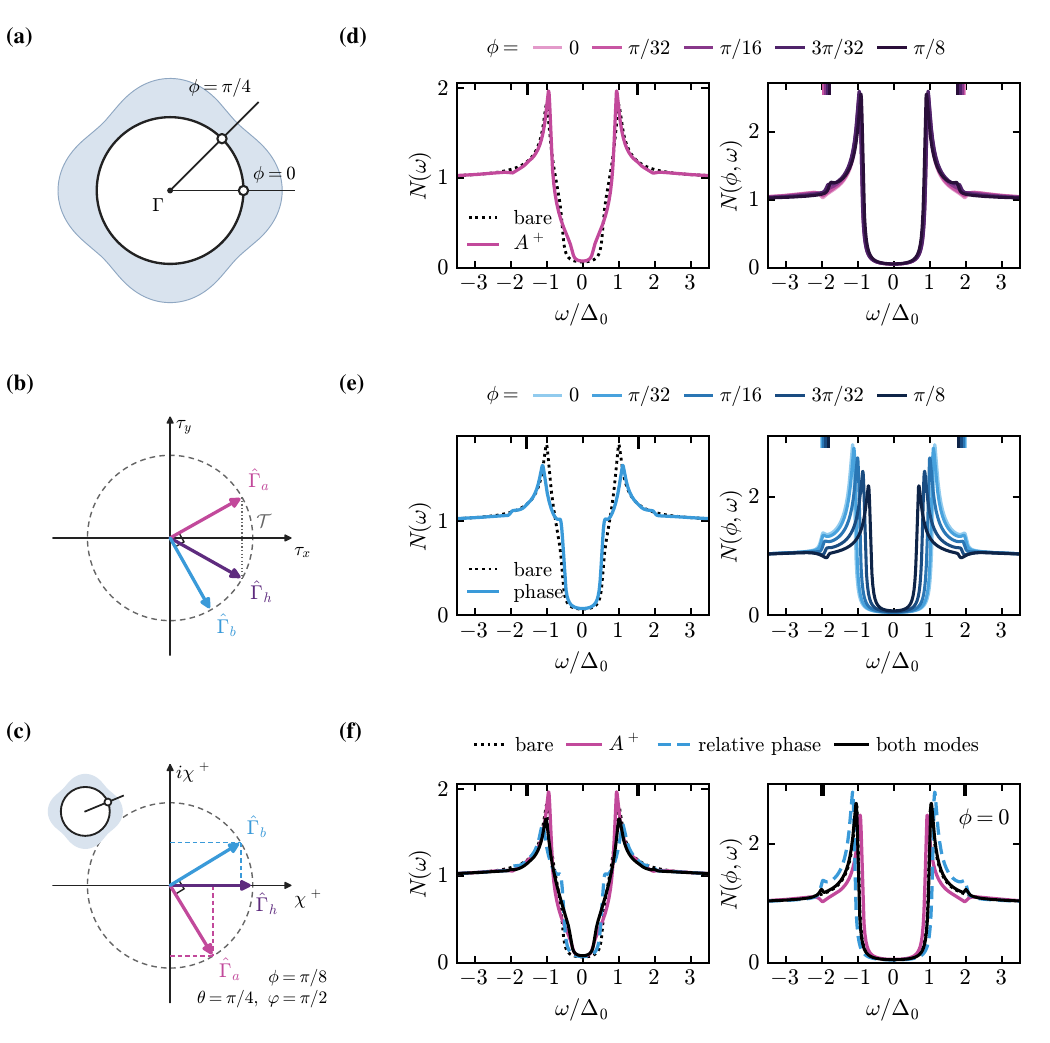}
  \caption{\textbf{Collective-mode vertices and their effect on the tunnelling DOS in the TRSB state} ($\theta=\pi/4$, $\varphi=\pi/2$).   \textbf{a}, Circular Fermi surface with the angle $\phi$ measured from the antinodal direction. The shaded region shows the anisotropic gap $\Delta_0|\chi^+(\phi)|$, largest at $\phi=0$ and smallest at $\phi=\pi/4$.   \textbf{b}, Higgs ($\hat\Gamma_h$), relative-amplitude ($\hat\Gamma_a$) and relative-phase ($\hat\Gamma_b$) vertices in the $(\tau_x,\tau_y)$ pseudospin plane. $\hat\Gamma_h$ and $\hat\Gamma_a$ are related by time reversal $\mathcal T$, and $\hat\Gamma_b\perp\hat\Gamma_a$.   \textbf{c}, The same vertices at $\phi=\pi/8$, decomposed along the local gap $\chi^+$ and the orthogonal direction $i\chi^+$.   \textbf{d}--\textbf{f}, Momentum-integrated DOS $N(\omega)$ (left) and angle-resolved DOS $N(\phi,\omega)$ (right) with (\textbf{d}) only the $A^+$ mode, (\textbf{e}) only the relative-phase mode and (\textbf{f}) both modes, compared with the bare spectrum (dotted). In \textbf{d} and \textbf{e} the right panels show $\phi$ between $0$ and $\pi/8$; in \textbf{f} the right panel shows $\phi=0$. Ticks at the top mark $\pm(\Omega_i+\Delta_{\min})$ (left) and $\pm(\Omega_i+\Delta_0|\chi^+(\phi)|)$ (right).}
  \label{fig:did_asy}
\end{figure}

\begin{figure}[p]
  \centering
  \includegraphics[scale=.8]{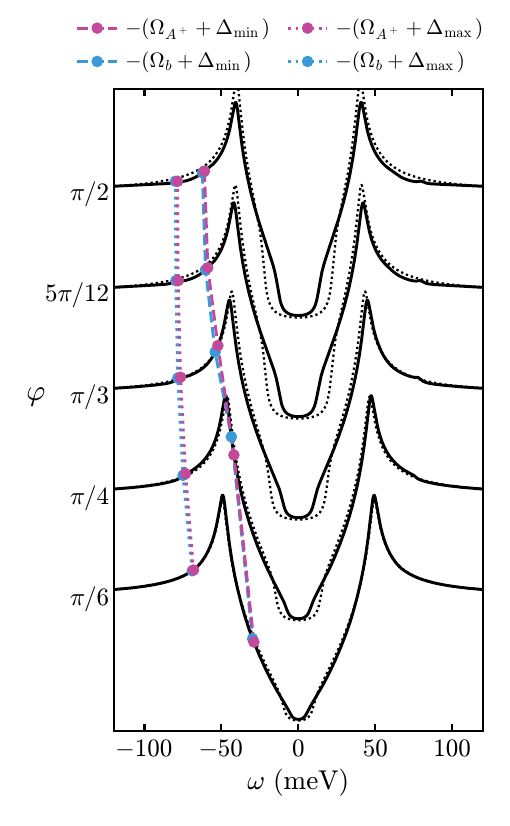}
  \caption{\textbf{Momentum-integrated tunnelling DOS across the chiral phase.} $N(\omega)$, in units of the normal-state value $N_{\mathrm n}$, versus bias energy $\omega$ for relative phase $\varphi$ between $\pi/6$ and $\pi/2$ ($\theta$ between $31^\circ$ and $45^\circ$). Solid lines include both the $A^+$ and the relative-phase mode; dotted lines show the bare spectrum. Curves are offset vertically for clarity. Markers indicate $-(\Omega_i+\Delta_{\min})$ (dashed) and $-(\Omega_i+\Delta_{\max})$ (dotted) for the $A^+$ mode (pink) and the relative-phase mode (blue). Over this range $\Delta_{\min}$ goes from $22.5$ to $9.9$~meV, $\Delta_{\max}$ from $40$ to $49$~meV, $\Omega_{A^+}$ from $38.6$ to $19.0$~meV and $\Omega_b$ from $39.9$ to $19.8$~meV. Parameters: $\Delta_0=40$~meV, $t=30$~meV, $\mathcal{K}=0.136$, $\hbar v_F=50$~meV$\,a_{\mathrm{lat}}$ ($a_{\mathrm{lat}}$ is the length unit used for momenta), Dynes and self-energy broadening $\Gamma_0=\gamma_\Sigma=2$~meV.}
  \label{fig:waterfall_int}
\end{figure}

\begin{figure}[p]
  \centering
  \includegraphics[scale=.8]{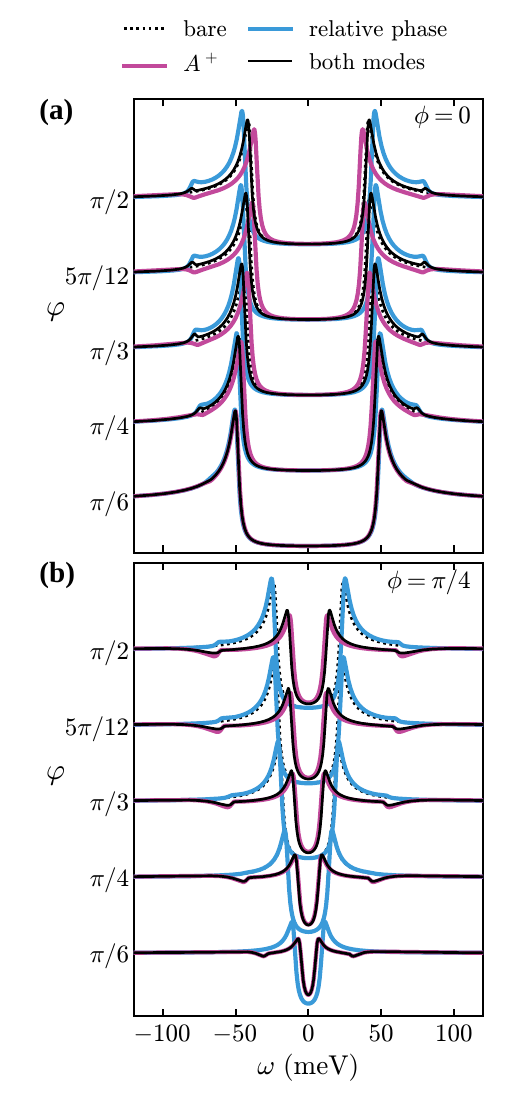}
  \caption{\textbf{Angle-resolved tunnelling DOS.} Angle-resolved, radially integrated DOS $N(\phi,\omega)$, in units of the normal-state value, versus bias energy $\omega$ along (\textbf{a}) the antinodal direction $\phi=0$ and (\textbf{b}) the nodal direction $\phi=\pi/4$, for relative phase $\varphi$ between $\pi/6$ and $\pi/2$. Each spectrum is shown for the bare superconductor (dotted), with only the $A^+$ mode (pink), with only the relative-phase mode (blue) and with both modes (black). Curves are offset vertically for clarity. Parameters as in \Fig{fig:waterfall_int}.}
  \label{fig:waterfall_k}
\end{figure}

\begin{figure}[p]
  \centering
  \includegraphics[width=0.92\textwidth]{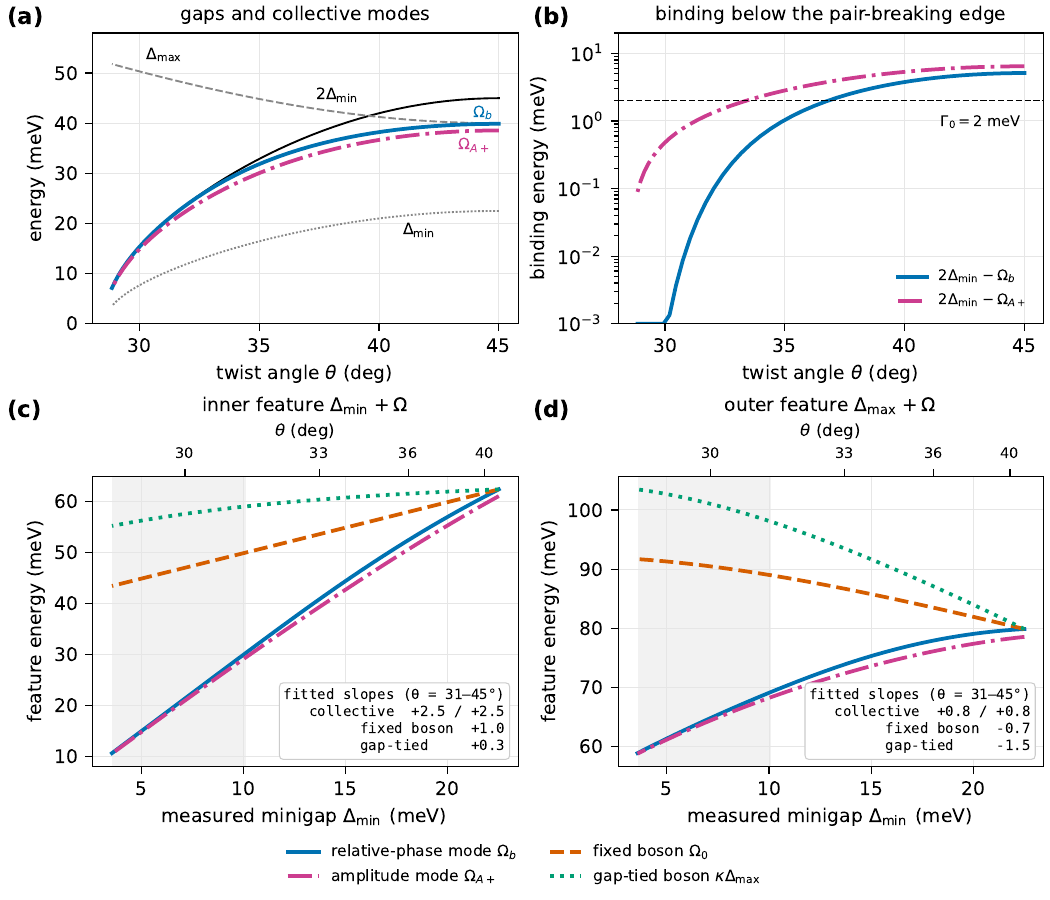}
  \caption{\textbf{Twist-angle test for intrinsic collective modes.}   \textbf{a}, Minigap $\Delta_{\min}$, maximum gap $\Delta_{\max}$, pair-breaking edge $2\Delta_{\min}$, and the energies of the relative-phase ($\Omega_b$) and hybrid amplitude ($\Omega_{A+}$) modes versus twist angle. \textbf{b}, Binding energy $2\Delta_{\min}-\Omega_i$ (log scale); the dashed line marks $\Gamma_0=2$~meV. \textbf{c}, Inner and (\textbf{d}) outer resonance energies against the minigap $\Delta_{\min}$ (bottom axis; the corresponding twist angle is shown on top), for the two collective modes, a fixed-energy boson $\Omega_0=\Omega_b(45^\circ)$, and a boson tied to the intrinsic gap, $\Omega=\kappa\Delta_{\max}$ with $\kappa=\Omega_b(45^\circ)/\Delta_{\max}(45^\circ)$. Slopes are linear fits over $31^\circ\le\theta\le45^\circ$ ($\varphi\in[\pi/6,\pi/2]$, as in \Fig{fig:waterfall_int}); the shaded region lies outside this range. Parameters as in \Fig{fig:waterfall_int}.}
  \label{fig:discriminant}
\end{figure}

\begin{figure}[p]
  \centering
  \includegraphics[scale=.8]{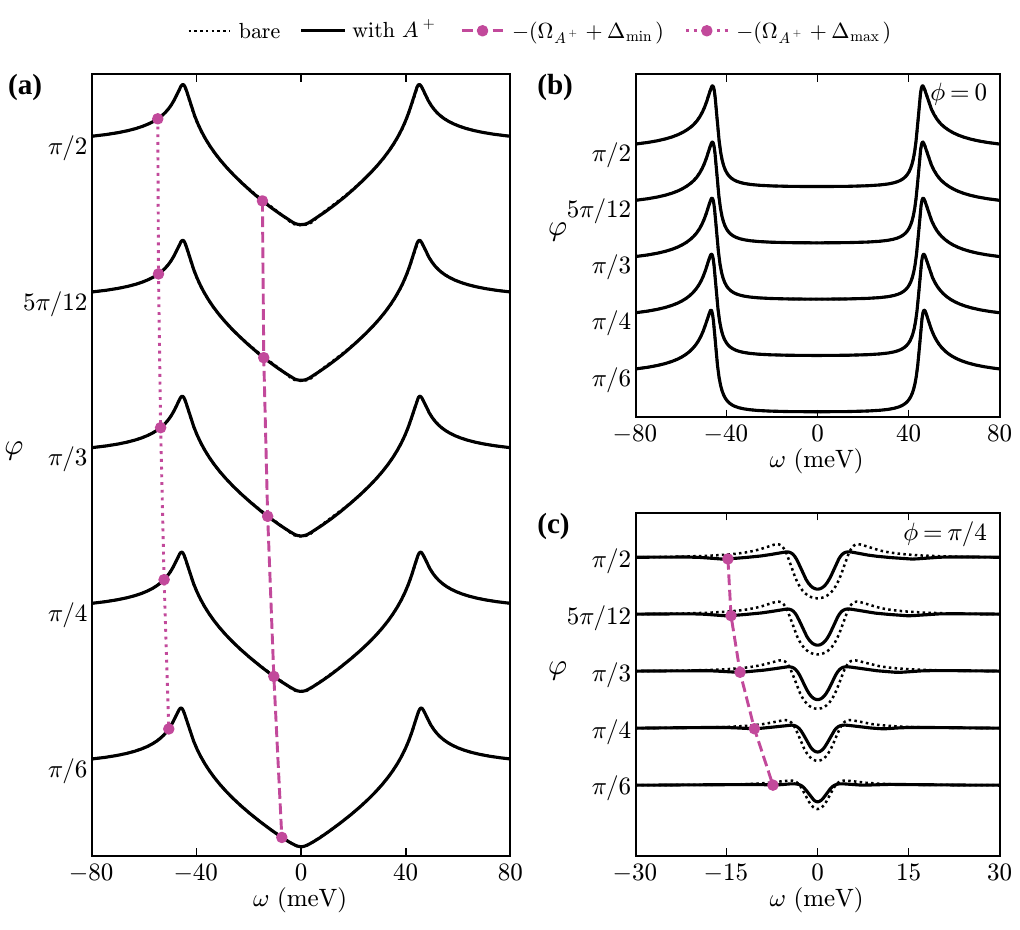}
 \caption{\textbf{Tunnelling DOS for the experimental parameters.} (\textbf{a}) Momentum-integrated DOS $N(\omega)$ and angle-resolved, radially integrated DOS $N(\phi,\omega)$ along (\textbf{b}) the antinodal direction $\phi=0$ and (\textbf{c}) the nodal direction $\phi=\pi/4$, in units of the normal-state value $N_{\mathrm n}$, versus bias energy $\omega$ for relative phase $\varphi$ between $\pi/6$ and $\pi/2$ ($\theta$ between $40.5^\circ$ and $45^\circ$). For these parameters the relative-phase mode lies in the continuum, so only the $A^+$ mode is retained. Solid lines include the coupling to the $A^+$ mode; dotted lines show the bare spectrum. Curves are offset vertically for clarity. Pink markers indicate $-(\Omega_{A^+}+\Delta_{\min})$ (dashed) and $-(\Omega_{A^+}+\Delta_{\max})$ (dotted) in (a), and $-(\Omega_{A^+}+\Delta_{\min})$ in (c), where $\Delta_{\min}$ is the local gap at $\phi=\pi/4$. From $\varphi=\pi/2$ to $\pi/6$, $\Delta_{\min}$ goes from $5.0$ to $2.5$~meV, $\Delta_{\max}$ from $45.0$ to $45.7$~meV and $\Omega_{A^+}$ from $9.8$ to $4.9$~meV. Parameters: $\Delta_1=45$~meV, $t=15$~meV, $\mathcal{K}=0.045$, $\hbar v_F=50$~meV$\,a_{\mathrm{lat}}$, $\Gamma_0=\gamma_\Sigma=2$~meV.}
  \label{fig:t15_combined}
\end{figure}

\clearpage
\setcounter{equation}{0}\setcounter{figure}{0}\setcounter{table}{0}
\renewcommand{\theequation}{S\arabic{equation}}
\renewcommand{\thefigure}{\arabic{figure}}
\renewcommand{\thetable}{\arabic{table}}
\renewcommand{\figurename}{Supplementary Figure}
\renewcommand{\tablename}{Supplementary Table}
\renewcommand{\theHequation}{S\arabic{equation}}
\renewcommand{\theHfigure}{S\arabic{figure}}
\renewcommand{\theHtable}{S\arabic{table}}
\begin{center}
{\Large\textbf{Supplementary Information}}\\[6pt]
{\large Twist-tunable clapping modes and their tunnelling signatures in bilayer cuprates}
\end{center}

\section*{SUPPLEMENTARY NOTE 1}
\subsection*{Model for twisted bilayer cuprates}

The Bogoliubov--de Gennes (BdG) Hamiltonian of the two layers reads
\begin{equation}
h_{\mathbf k}=\begin{pmatrix}
\xi_{\mathbf k} & \Delta_{\mathbf k,1} & t & 0 \\
\Delta_{\mathbf k,1}^* & -\xi_{\mathbf k} & 0 & -t \\
t & 0 & \xi_{\mathbf k} & \Delta_{\mathbf k,2} \\
0 & -t & \Delta_{\mathbf k,2}^* & -\xi_{\mathbf k}
\end{pmatrix},
\label{eq:H0}
\end{equation}
written in the four-component Nambu spinor $\Psi_{\mathbf k}=(c_{\mathbf k\uparrow1},c^\dagger_{-\mathbf k\downarrow1},c_{\mathbf k\uparrow2},c^\dagger_{-\mathbf k\downarrow2})^T$, so that $\mathcal H=\sum_{\mathbf k}\Psi^\dagger_{\mathbf k}h_{\mathbf k}\Psi_{\mathbf k}$ up to a constant. Here $\xi_{\mathbf k}$ is the band dispersion, taken parabolic and equal in the two layers, $t$ is the interlayer tunnelling, and $\Delta_{\mathbf k,a}$ is the intrinsic $d$-wave gap of layer $a=1,2$. Integrating out the electrons of the bottom layer~\cite{PhysRevLett.132.017002} gives the effective Hamiltonian of the top layer, equation~\eqref{eq:Heff_main} of the main text, with the form factors of equation~\eqref{eq:formfactors} and the proximity-induced gap $\Delta_2\simeq t^2/\Delta_1$~\cite{Can}.

\subsection*{Self-consistency constraints on the induced order parameter}
\label{sec:selfconsistency}

In this section we derive the symmetry constraints imposed by the nonlinear gap equation on the proximity-induced superconducting state of the top layer. Our analysis is based on Ref.~\cite{PhysRevB.53.3598} for mixed-symmetry superconductors, generalized here to the twisted $(d_{x^2-y^2},d_{xy})$ sector generated by the relative rotation between the two layers.

We start from the effective Hamiltonian of equation~\eqref{eq:Heff_main} of the main text, and denote its off-diagonal element by $\Psi(\mathbf k)$, so that $E_{\mathbf k}=\sqrt{\tilde\xi_{\mathbf k}^2+|\Psi(\mathbf k)|^2}$. The order parameter of the top layer transforms as $f_{d_{x^2-y^2}}(\mathbf k)=\cos2\phi_{\mathbf k}$, while that of the bottom layer is rotated by the twist angle $\theta$, $f_{2,\mathbf k}=\cos(2\phi_{\mathbf k}-2\theta)$. Using trigonometric identities, $f_{2,\mathbf k}=f_{d_{x^2-y^2}}\cos2\theta+f_{d_{xy}}\sin2\theta$, with $f_{d_{xy}}(\mathbf k)=\sin2\phi_{\mathbf k}$. A natural ansatz for the induced pairing amplitude therefore reads
\begin{equation}
\Psi(\mathbf k)=\Delta_1f_{d_{x^2-y^2}}+e^{i\varphi}\Delta_2\left(f_{d_{x^2-y^2}}\cos2\theta+f_{d_{xy}}\sin2\theta\right).
\label{eq:naiveansatz}
\end{equation}
Equation~\eqref{eq:naiveansatz} assumes that the same global phase factor $e^{i\varphi}$ multiplies both irreducible representations generated by the twist. As we show below, this structure is not generally compatible with the self-consistent gap equation. To analyze the problem, we introduce the most general complex order parameter in the $(d_{x^2-y^2},d_{xy})$ sector,
\begin{equation}
\Psi(\mathbf k)=A_{d}f_{d_{x^2-y^2}}+A_{xy}f_{d_{xy}},\qquad
A_d=\Delta_d'+i\Delta_d'',\quad A_{xy}=\Delta_{xy}'+i\Delta_{xy}''.
\label{eq:generalpsi}
\end{equation}
The corresponding quasiparticle energy contains
\begin{equation}
|\Psi(\mathbf k)|^2=|A_d|^2f_{d_{x^2-y^2}}^2+|A_{xy}|^2f_{d_{xy}}^2+2C\,f_{d_{x^2-y^2}}f_{d_{xy}},
\qquad C=\mathrm{Re}(A_dA_{xy}^*)=\Delta_d'\Delta_{xy}'+\Delta_d''\Delta_{xy}''.
\label{eq:modpsi}
\end{equation}
The self-consistency equation takes the standard form
\begin{equation}
\Psi(\mathbf k)=-\sum_{\mathbf k'}V(\mathbf k,\mathbf k')\frac{\Psi(\mathbf k')}{2E_{\mathbf k'}}\tanh\frac{E_{\mathbf k'}}{2T}.
\label{eq:gapequation}
\end{equation}
Projecting the interaction onto the two pairing channels, $V(\mathbf k,\mathbf k')=V_df_{d_{x^2-y^2}}(\mathbf k)f_{d_{x^2-y^2}}(\mathbf k')+V_{xy}f_{d_{xy}}(\mathbf k)f_{d_{xy}}(\mathbf k')$, one obtains separate projected gap equations for the two irreducible representations. For example, projection onto the $d_{x^2-y^2}$ channel yields
\begin{equation}
A_d=-V_d\sum_{\mathbf k}f_{d_{x^2-y^2}}\frac{A_df_{d_{x^2-y^2}}+A_{xy}f_{d_{xy}}}{2E_{\mathbf k}}\tanh\frac{E_{\mathbf k}}{2T}.
\label{eq:dprojection}
\end{equation}
The mixed harmonic $f_{d_{x^2-y^2}}f_{d_{xy}}$ changes sign under $k_x\leftrightarrow k_y$, while $f_{d_{x^2-y^2}}^2$ and $f_{d_{xy}}^2$ remain invariant. As shown by Musaelian \textit{et al.}~\cite{PhysRevB.53.3598}, the self-consistent solution of the coupled gap equations then satisfies $C=0$,
\begin{equation}
\Delta_d'\Delta_{xy}'+\Delta_d''\Delta_{xy}''=0,
\label{eq:constraint}
\end{equation}
the direct analogue of the phase-locking condition derived in Ref.~\cite{PhysRevB.53.3598}. A solution with both components finite and satisfying Eq.~\eqref{eq:constraint} is the chiral configuration $\Delta_{xy}'=\Delta_d''=0$, $\Psi(\mathbf k)=\Delta_d'f_{d_{x^2-y^2}}+i\Delta_{xy}''f_{d_{xy}}$.

Expanding Eq.~\eqref{eq:naiveansatz} into real and imaginary parts gives
\begin{equation}
\Psi(\mathbf k)=\left[\Delta_1+\Delta_2\cos\varphi\cos2\theta\right]f_{d_{x^2-y^2}}+\Delta_2\cos\varphi\sin2\theta\,f_{d_{xy}}
+i\Delta_2\sin\varphi\left(\cos2\theta\,f_{d_{x^2-y^2}}+\sin2\theta\,f_{d_{xy}}\right).
\label{eq:split}
\end{equation}
For generic values of $(\theta,\varphi)$ both irreducible representations appear simultaneously in the real and imaginary sectors, $\Delta_{xy}'\neq0$ and $\Delta_d''\neq0$, which violates Eq.~\eqref{eq:constraint}. The self-consistency equation therefore rejects the global-phase ansatz of Eq.~\eqref{eq:naiveansatz} except at special symmetry points. Retaining in each irreducible representation the component compatible with Eq.~\eqref{eq:constraint}, the two representations separate into orthogonal phase sectors, leading to
\begin{equation}
\Psi(\mathbf k)=\left[\Delta_1+\Delta_2\cos2\theta\cos\varphi\right]f_{d_{x^2-y^2}}+i\,\Delta_2\sin2\theta\sin\varphi\,f_{d_{xy}},
\label{eq:finalresult}
\end{equation}
i.e.\ equation~\eqref{eq:OP_final} of the main text with $r=\Delta_2/\Delta_1$, which coincides with Eq.~\eqref{eq:naiveansatz} at $\theta=\pi/4$, $\varphi=\pi/2$. The rotated $d_{x^2-y^2}$ component coherently enhances the gap of the top layer, while the induced $id_{xy}$ component gaps the nodal quasiparticles and drives the system into a fully gapped chiral phase. 

\paragraph{Scope of the ansatz.}
Equation~\eqref{eq:finalresult} is obtained by imposing the symmetry constraint~\eqref{eq:constraint} on the induced component, rather than by solving the full bilayer gap equations self-consistently. The proximity-induced term is inherited from the bottom layer, whose order parameter is not itself determined by the top-layer gap equation, so the constraint should be understood as selecting the chiral structure that is compatible with a self-consistent top-layer solution. Equations~\eqref{eq:naiveansatz} and \eqref{eq:finalresult} coincide at $\theta=\pi/4$, $\varphi=\pi/2$, and differ by terms of order $r\cos\varphi\sin2\theta$ in the real $d_{xy}$ and $r\sin\varphi\cos2\theta$ in the imaginary $d_{x^2-y^2}$ channel away from this point. A fully self-consistent bilayer calculation would quantify these corrections; we expect them to shift the mode energies but not their pinning to $2\Delta_{\min}$, on which the twist-angle test of the main text relies.

\section*{SUPPLEMENTARY NOTE 2}
\subsection*{Collective modes through the phase diagram}

Besides the Anderson--Bogoliubov--Goldstone mode, the order parameter supports three massive collective modes: the Higgs mode and the relative-amplitude and relative-phase modes. We compute their propagators following Ref.~\cite{Poniatowski2022}, starting from the action of equation~\eqref{eq:action} of the main text written in the chiral basis of equation~\eqref{eq:chi}. We write the order parameter as $\Delta_{\mathbf k}=\Delta^+\chi^+_{\mathbf k}+\Delta^-\chi^-_{\mathbf k}$ and take the $\chi^+$ channel as the condensed one, so that at the saddle point $\Delta^+=\Delta_0\equiv\Delta_1$ and $\Delta^-=0$. Following Ref.~\cite{Poniatowski2022}, we take the pairing interaction diagonal in this basis, with couplings $g_\pm>0$. The gap equation of the condensed channel at $T=0$ reads
\begin{equation}
\frac{1}{g_+}=\int_{\mathbf k}\frac{|\chi^+_{\mathbf k}|^2}{2E_{\mathbf k}},
\qquad E_{\mathbf k}=\sqrt{\xi_{\mathbf k}^2+\Delta_0^2|\chi^+_{\mathbf k}|^2},
\label{eq:gap_eq}
\end{equation}
and is used below to eliminate $g_+$. A similar decomposition for a chiral order parameter with non-orthogonal components was discussed in Ref.~\cite{JURECKA1999304}.

The inverse Gor'kov Green's function in this basis is
\begin{equation}
G^{-1}(\mathbf k,\tau)=-\partial_\tau-\xi_{\mathbf k}\tau_z-\left[\tau_+\left(\Delta^+\chi^+_{\mathbf k}+\Delta^-\chi^-_{\mathbf k}\right)+\tau_-\left(\Delta^{+*}\chi^-_{\mathbf k}+\Delta^{-*}\chi^+_{\mathbf k}\right)\right],
\end{equation}
where $\tau_{x,y,z}$ are Pauli matrices in Nambu space and $\tau_\pm=(\tau_x\pm i\tau_y)/2$. We parametrize the fluctuations around the saddle point as
\begin{equation}
\Delta^+(\tau)=\left[\Delta_0+h(\tau)\right]e^{i\theta^+(\tau)},\qquad
\Delta^-(\tau)=\left[a(\tau)+i\,b(\tau)\right]e^{i\theta^+(\tau)},
\end{equation}
where $h$ is the amplitude (Higgs) fluctuation of the condensed channel and $\theta^+$ its phase. Since $\Delta^-$ vanishes at the saddle point, its fluctuation is fully described by the two real fields $a$ and $b$, its components in phase and in quadrature with the condensate, i.e.\ the relative-amplitude and relative-phase fluctuations. The overall phase $\theta^+$ is removed by the gauge transformation $\psi\to e^{i\theta^+\tau_z/2}\psi$, which generates the term $-\tfrac{i}{2}\partial_\tau\theta^+\tau_z$ coupling to the Goldstone mode. In Matsubara and momentum space we then write $G^{-1}=G_0^{-1}-\hat V$, with
\begin{align}
G_0^{-1}(\mathbf k,i\omega_n)&=i\omega_n-\xi_{\mathbf k}\tau_z-\Delta_0\hat\Gamma_{h,\mathbf k},\\
\hat V(\mathbf k,\mathbf q,i\Omega_m)&=h_q\,\hat\Gamma_{h,\mathbf k}+a_q\,\hat\Gamma_{a,\mathbf k}+b_q\,\hat\Gamma_{b,\mathbf k}+\tfrac{1}{2}\Omega_m\theta^+_q\,\tau_z,
\end{align}
where $\hat\Gamma_{h,a,b}$ are the vertices of equation~\eqref{eq:vertices} of the main text. Integrating out the fermions,
\begin{equation}
S_{\mathrm{eff}}=\sum_{s,q}\frac{|\Delta^s_q|^2}{g_s}-\mathrm{Tr}\ln\left(-G_0^{-1}\right)+\sum_{n=1}^\infty\frac{1}{n}\mathrm{Tr}\left(G_0\hat V\right)^n,
\end{equation}
and the term with $n=2$ yields the Gaussian action of the collective modes.

At $T=0$ and $\mathbf q=0$ the bare inverse propagators, after continuation $i\Omega_m\to\Omega+i0^+$, read
\begin{align}
D_h^{-1}(\Omega)&=\int_{\mathbf k}\frac{1}{2E_{\mathbf k}}\,\frac{|\chi^+_{\mathbf k}|^2\left(4\Delta_0^2|\chi^+_{\mathbf k}|^2-\Omega^2\right)}{\Omega^2-4E_{\mathbf k}^2},
\label{eq:Dh}\\
D_a^{-1}(\Omega)&=\int_{\mathbf k}\frac{1}{2E_{\mathbf k}}\,\frac{2\Delta_0^2|\chi^+_{\mathbf k}|^4+\Delta_0^2\left[(\chi^+_{\mathbf k})^4+(\chi^-_{\mathbf k})^4\right]-|\chi^+_{\mathbf k}|^2\,\Omega^2}{\Omega^2-4E_{\mathbf k}^2},
\label{eq:Da}\\
D_b^{-1}(\Omega)&=\int_{\mathbf k}\frac{1}{2E_{\mathbf k}}\,\frac{2\Delta_0^2|\chi^+_{\mathbf k}|^4-\Delta_0^2\left[(\chi^+_{\mathbf k})^4+(\chi^-_{\mathbf k})^4\right]-|\chi^+_{\mathbf k}|^2\,\Omega^2}{\Omega^2-4E_{\mathbf k}^2},
\label{eq:RP}
\end{align}
where the gap equation~\eqref{eq:gap_eq} has been used in $D_h^{-1}$. The Higgs and relative-amplitude fluctuations are coupled by
\begin{equation}
\Pi^{ha}(\Omega)=\int_{\mathbf k}\frac{1}{2E_{\mathbf k}}\,\frac{\left[(\chi^+_{\mathbf k})^2+(\chi^-_{\mathbf k})^2\right]\left(\Omega^2-4\Delta_0^2|\chi^+_{\mathbf k}|^2\right)}{\Omega^2-4E_{\mathbf k}^2}.
\label{eq:Pha}
\end{equation}
Since $\chi^\pm_{\mathbf k}$ and $\chi^\mp_{\mathbf k}$ are related by the mirror $\phi_{\mathbf k}\to-\phi_{\mathbf k}$, the couplings $\Pi^{hb}$ and $\Pi^{ab}$, whose integrands contain $(\chi^+_{\mathbf k})^n-(\chi^-_{\mathbf k})^n$, vanish after the Fermi-surface integration. In Eq.~\eqref{eq:Pha} we have dropped the frequency-independent term $\int_{\mathbf k}\left[(\chi^+_{\mathbf k})^2+(\chi^-_{\mathbf k})^2\right]/2E_{\mathbf k}$, which we assume to be compensated by the interaction term that keeps $\Delta^-=0$ a saddle point, in the same way as the gap equation~\eqref{eq:gap_eq} removes the frequency-independent part of $D_h^{-1}$. The Gaussian action therefore separates into a phase and an amplitude sector,
\begin{equation}
S_{\mathrm{fluct}}=-\sum_q
\begin{pmatrix} h_{-q} & a_{-q} & b_{-q}\end{pmatrix}
\begin{pmatrix}
D_h^{-1} & -\tfrac12\Pi^{ha} & 0\\
-\tfrac12\Pi^{ha} & D_a^{-1} & 0\\
0 & 0 & D_b^{-1}
\end{pmatrix}
\begin{pmatrix} h_q\\ a_q\\ b_q\end{pmatrix}.
\end{equation}
The relative-phase mode is the zero of $D_b^{-1}$. The two amplitude modes $A^\pm$ are given by the eigenvalues of the $2\times2$ block,
\begin{equation}
D_{A^\pm}^{-1}=\frac{D_h^{-1}+D_a^{-1}}{2}\pm\frac12\sqrt{\left(D_h^{-1}-D_a^{-1}\right)^2+\left(\Pi^{ha}\right)^2},
\label{eq:DApm}
\end{equation}
and the corresponding eigenvector $(\cos\beta_{A^+},\sin\beta_{A^+})$ at the pole defines the mixing angle of equation~\eqref{eq:GammaAplus} of the main text. These are the propagators shown in Fig.~\ref{fig:modes} of the main text.

\section*{SUPPLEMENTARY NOTE 3}
\subsection*{Bare density of states}

\begin{figure}[t!]
\centering
\includegraphics[width=0.6\textwidth]{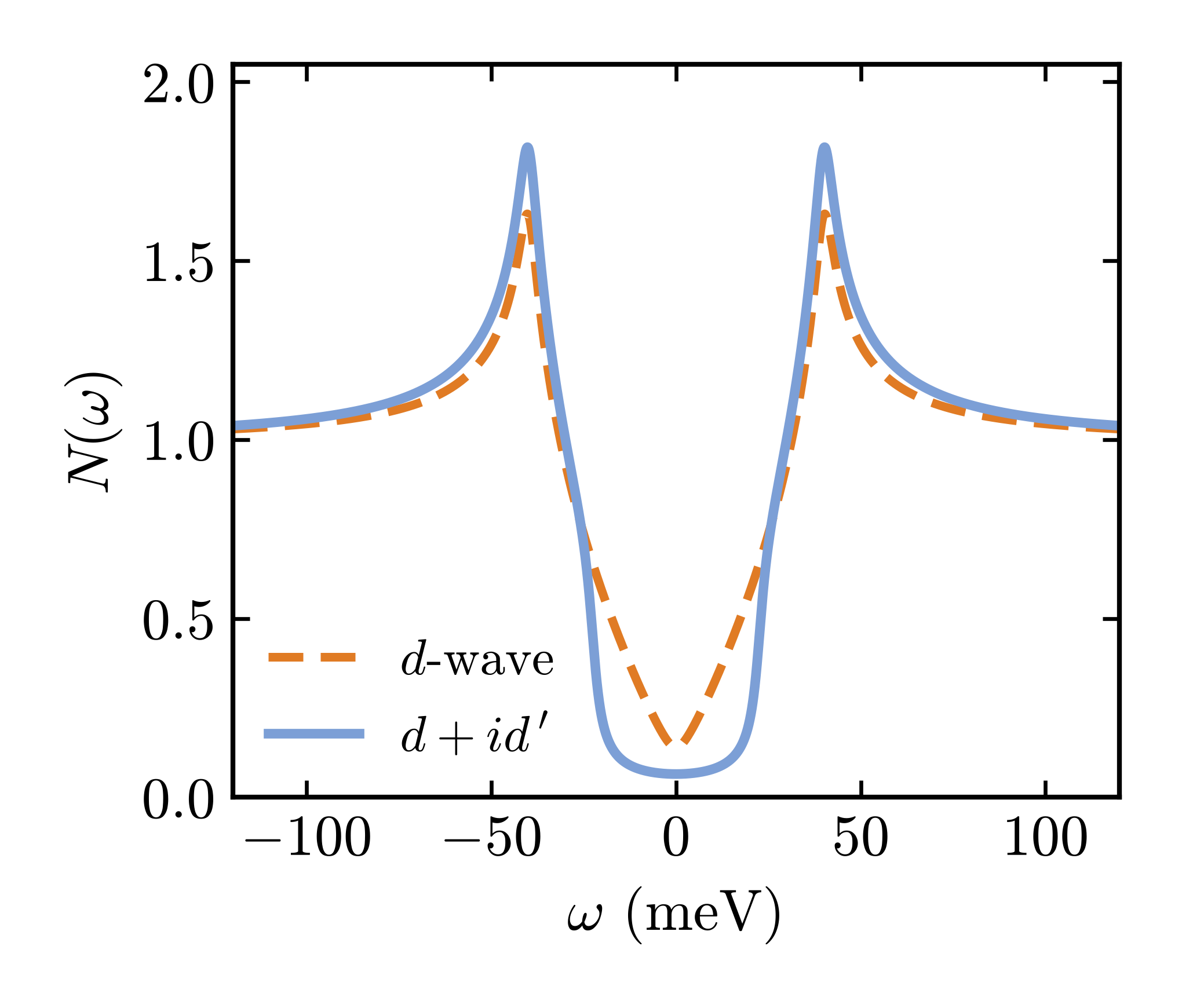}
\caption{Bare density of states of a single $d$-wave layer (dashed) and of the effective model of equation~\eqref{eq:OP_final} of the main text at $\theta=\pi/4$, $\varphi=\pi/2$ (solid). Parameters: $\Delta_1=40$~meV, $t=30$~meV, $\Delta_2=t^2/\Delta_1$, $\Gamma_0=2$~meV.}
\label{sfig:DOS}
\end{figure}

Supplementary Fig.~\ref{sfig:DOS} shows how the gap opens in the TRSB state, using equation~\eqref{eq:Ak} of the main text without self-energy. In contrast to the V-shaped spectrum of the $d$-wave state, the spectrum is fully gapped, with a shoulder at $\Delta_{\min}=\Delta_0\min_\phi|\chi^+(\phi)|$ and coherence peaks at $\Delta_{\max}=\Delta_0\max_\phi|\chi^+(\phi)|$.

\subsection*{Pole approximation of the bosonic propagator}

After continuation $i\Omega_m\to\Omega+i0^+$, each inverse propagator of Supplementary Note~2 has two features. The first is a zero at a real frequency $\Omega_i<2\Delta_{\min}$, which corresponds to a sharp, undamped collective mode: below the pair-breaking threshold the inverse propagator has no imaginary part. The second is a branch cut starting at $2\Delta_{\min}$, which describes the decay of the mode into two quasiparticles and gives a smooth, incoherent background. We retain only the pole~\cite{PhysRevB.49.740}. Near $\Omega_i$ the inverse propagator vanishes linearly in $\Omega^2$,
\begin{equation}
D_i^{-1}(\Omega)\simeq\frac{\Omega^2-\Omega_i^2}{Z_i},
\qquad
Z_i^{-1}=\left.\frac{\partial D_i^{-1}}{\partial(\Omega^2)}\right|_{\Omega=\Omega_i},
\label{eq:Zb}
\end{equation}
which gives the frequency dependence of equation~\eqref{eq:bosono_prop} of the main text. The approximation is controlled when the mode lies well inside the gap, where the pole and the continuum are separated in frequency and the contribution of the continuum to the self-energy is suppressed by the energy denominators. The mode frequency $\Omega_i(\varphi)$ is obtained numerically as the root of $\operatorname{Re}D_i^{-1}(\Omega,\varphi)=0$ below $2\Delta_{\min}(\varphi)$, using Brent's method, with $D_b^{-1}$ from Eq.~\eqref{eq:RP} and $D_{A^+}^{-1}$ from Eq.~\eqref{eq:DApm}. The residue $Z_i(\varphi)$ follows from Eq.~\eqref{eq:Zb}, evaluated with a finite-difference derivative at the pole.

\paragraph{Momentum profile.}
Equations~\eqref{eq:Dh}--\eqref{eq:Pha} are written at zero momentum transfer. At small $\mathbf q$ the bubbles acquire a gradient expansion~\cite{thz_near}. As is common in the literature on electron--boson coupling in unconventional superconductors, we assume that the frequency and momentum dependence of the propagator factorize, as in equation~\eqref{eq:bosono_prop} of the main text. Eschrig and Norman~\cite{PhysRevLett.85.3261} used the same strategy for the spin resonance in the cuprates, separating a momentum envelope from the energy distribution of the resonance. To fix the correlation length $\ell_i$ in equation~\eqref{eq:bosono_prop} of the main text, we assume that the leading momentum dependence of the inverse propagator has the same form as for the Higgs mode in two dimensions~\cite{thz_near},
\begin{equation}
D_i^{-1}(\nu,\mathbf q)\propto\Omega_i^2+\tfrac12v_F^2q^2-\nu^2 .
\label{eq:D_full}
\end{equation}
The correlation length is a static quantity, and setting $\nu=0$ gives
\begin{equation}
D_i(0,\mathbf q)\propto\frac{1}{\Omega_i^2}\,\frac{1}{1+\dfrac{v_F^2}{2\Omega_i^2}q^2},
\qquad\text{so that}\qquad
\ell_i=\frac{v_F}{\sqrt2\,\Omega_i}.
\label{eq:Fq}
\end{equation}

\subsection*{Self-energy in the forward-scattering approximation}

Here we evaluate the one-loop self-energy, equation~\eqref{eq:sigma} of the main text, within the local approximation described in Methods.

\paragraph{Vertices.}
We write each vertex as $\hat\Gamma_i=\tau_+v_i+\tau_-v_i^*$. From equations~\eqref{eq:vertices} and \eqref{eq:GammaAplus} of the main text,
\begin{equation}
v_b=i\chi^-,\qquad v_{A^+}=\cos\beta_{A^+}\,\chi^++\sin\beta_{A^+}\,\chi^-.
\end{equation}

\paragraph{Momentum integral.}
We take $\mathbf k$ on the Fermi surface at angle $\phi$ and write $\mathbf q=q(\cos\theta_q,\sin\theta_q)$, with $\theta_q$ measured from the Fermi velocity $\mathbf v_F(\phi)$. Linearizing the dispersion,
\begin{equation}
\xi_{\mathbf k-\mathbf q}\simeq-v_Fq\cos\theta_q,
\qquad
q\,dq=\frac{|\xi|}{v_F^2\cos^2\theta_q}\,d\xi.
\label{eq:lin_disp}
\end{equation}
Only the component of $\mathbf q$ along $\mathbf v_F$ changes the quasiparticle energy, while the full $|\mathbf q|=|\xi|/(v_F|\cos\theta_q|)$ enters $F_i$. With $E=\sqrt{\xi^2+d^2}$, where $d=\Delta_0|\chi^+(\phi)|$ is the local gap, one has $|\xi|d\xi=E\,dE$, and the angular integral can be done exactly,
\begin{equation}
\begin{aligned}
\int_0^{2\pi}\frac{d\theta_q}{\cos^2\theta_q}\,F_i\!\left(\frac{\sqrt{E^2-d^2}}{v_F|\cos\theta_q|}\right)
&=\int_0^{2\pi}\frac{d\theta_q}{\cos^2\theta_q+\ell_i^2(E^2-d^2)/v_F^2}\\
&=\frac{4\pi\,\Omega_i^2}{\sqrt{(E^2-d^2)(E^2-d^2+2\Omega_i^2)}}.
\end{aligned}
\end{equation}
The divergence of $1/\cos^2\theta_q$ for $\mathbf q\perp\mathbf v_F$ is thus cut off by $F_i$. The momentum integral produces the coupling $U_i$ of equation~\eqref{eq:Ui} of the main text, multiplied by the spectral weight
\begin{equation}
\rho_i(E)=\frac{1}{\sqrt{(E^2-d^2)(E^2-d^2+2\Omega_i^2)}},
\label{eq:rho}
\end{equation}
The weight $E\rho_i(E)$ is the BCS density of states $E/\sqrt{E^2-d^2}$, multiplied by the factor $1/\sqrt{E^2-d^2+2\Omega_i^2}$ that results from the momentum profile $F_i(\mathbf q)$ of the mode. This factor is close to $1/\sqrt2\,\Omega_i$ near the threshold $E\simeq d$ and decays as $1/E$ at high energy, so that the exchanged boson only scatters quasiparticles within an energy window of order $\Omega_i$ above the gap.

The overall prefactor follows from the $\mathbf q$-integral, $(2\pi)^{-2}\,(Z_i/2\Omega_i)\,4\pi\Omega_i^2/v_F^2=2U_i$, and from the coherence factors $u^2,v^2=\tfrac12(1\pm\xi/E)$: their part odd in $\xi$ cancels between $\pm\xi$, leaving a factor $\tfrac12$. The two kernels below therefore carry exactly the prefactor $U_i$ of equation~\eqref{eq:Ui} of the main text.
\paragraph{Self-energy.}
Performing the Matsubara sum at $T=0$, the terms odd in $\xi$ cancel, and the self-energy takes the form
\begin{equation}
\Sigma_{11}=\Sigma_{22}=|v_i|^2\,N_i(\phi,\omega),\qquad
\Sigma_{12}=\Delta_0\,v_i^2\,\chi^-\,L_i(\phi,\omega),\qquad
\Sigma_{21}=\Delta_0\,(v_i^*)^2\,\chi^+\,L_i(\phi,\omega),
\label{eq:Sigma_SM}
\end{equation}
summed over the modes $i=b,A^+$, with the scalar kernels
\begin{align}
N_i(\phi,\omega)&=U_i\int_d^{E_c}dE\,E\,\rho_i(E)\left[\frac{1}{z-\Omega_i-E}+\frac{1}{z+\Omega_i+E}\right],\\
L_i(\phi,\omega)&=U_i\int_d^{E_c}dE\,\rho_i(E)\left[\frac{1}{z-\Omega_i-E}-\frac{1}{z+\Omega_i+E}\right],
\end{align}
where $z=\omega+i\gamma_\Sigma$, $\gamma_\Sigma$ is a small broadening and $E_c$ a high-energy cutoff. The first term in each bracket describes the emission of a boson by a quasiparticle above the gap, the second by a hole below it. The real and imaginary parts are obtained from the same spectral integral and are therefore Kramers--Kronig consistent. The Fermi-surface angle $\phi$ enters only as a parameter, through the local gap $d$ and the vertices, and not as an integration variable.

The self-energies of the two modes are added before evaluating the density of states, equation~\eqref{eq:Ak} of the main text. The branch of the square root is chosen such that $N(\phi,\omega)>0$ and the integrand tends to $1$ for $|\omega|\to\infty$. In addition to the parameters listed in the main text, we use the cutoff $E_c=4$~eV.

\subsection*{Dependence on the interlayer tunnelling}

\begin{figure}
\centering
\includegraphics[width=\textwidth]{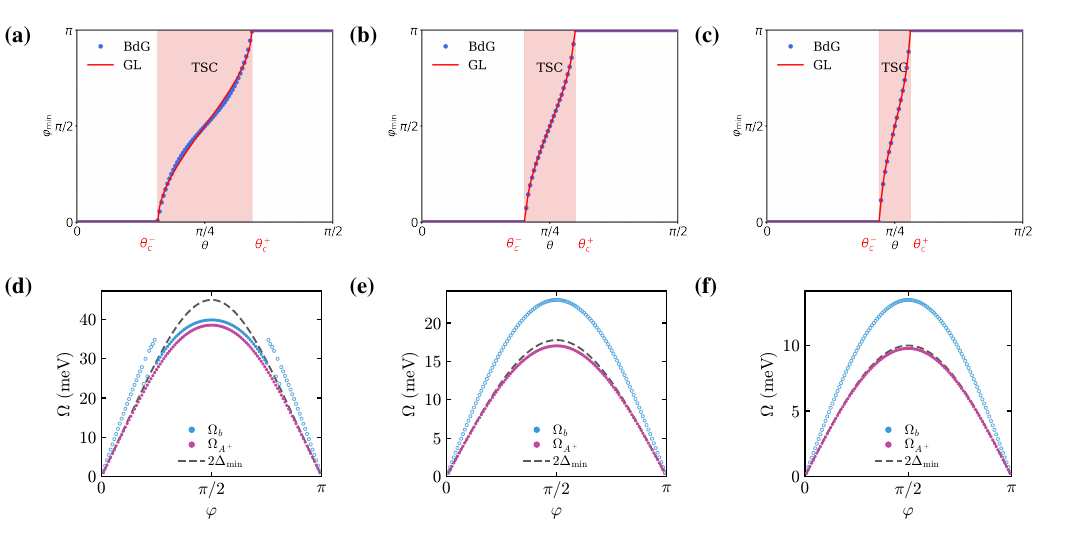}
\caption{Phase diagram and collective modes for different interlayer tunnelling. (a)--(c) Relative phase $\varphi_{\min}$ that minimizes the Ginzburg--Landau (line) and the BdG (symbols) free energy as a function of the twist angle $\theta$, for the effective model of equation~\eqref{eq:Heff_main} of the main text with $\Delta_2=t^2/\Delta_1$ and (a) $\Delta_1=40$~meV, $t=30$~meV, $\mathcal K=0.136$, $\theta_c^-=0.16\pi$, $\theta_c^+=0.34\pi$; (b) $\Delta_1=45$~meV, $t=20$~meV, $\mathcal K=0.073$, $\theta_c^-=0.20\pi$, $\theta_c^+=0.30\pi$; (c) $\Delta_1=45$~meV, $t=15$~meV, $\mathcal K=0.045$, $\theta_c^-=0.22\pi$, $\theta_c^+=0.28\pi$. The shaded region marks the TRSB phase. (d)--(f) Energies of the relative-phase mode $\Omega_b$ (blue) and of the amplitude mode $\Omega_{A^+}$ (pink) as a function of $\varphi$, obtained from $\operatorname{Re}D_b^{-1}=0$ and $\operatorname{Re}D_{A^+}^{-1}=0$, Eqs.~\eqref{eq:RP} and \eqref{eq:DApm}, for the parameters of (a)--(c). Filled symbols lie below the two-particle threshold $2\Delta_{\min}(\varphi)$ (dashed line) and correspond to sharp modes; open symbols lie inside the continuum. For $t=30$~meV both modes are sharp over a finite window around $\varphi=\pi/2$. For smaller $t$ the relative-phase mode moves into the continuum, while the $A^+$ mode remains just below $2\Delta_{\min}$.}

\label{sfig:many_t}
\end{figure}

The interlayer tunnelling $t$ sets the size of the proximity-induced gap, $\Delta_2\simeq t^2/\Delta_1$, and therefore the width of the TRSB window and the energies of the collective modes. The value $t=30$~meV used in the main text is the one adopted, together with $\Delta_1=40$~meV, in the continuum model of Ref.~\cite{Can}. Note that $t/\Delta_1=0.75$ is not small, so the integration over the bottom layer (Supplementary Note~1), which is perturbative in $t/\Delta_1$, should be regarded as a controlled description only for the smaller values of $t$ considered here. We also consider $t=20$~meV and $t=15$~meV. Supplementary Fig.~\ref{sfig:many_t} shows the phase diagram and the energies of the relative-phase and $A^+$ modes for the three values of $t$. For $t=30$~meV both modes lie below $2\Delta_{\min}$ over a finite window of $\varphi$ around $\pi/2$. As $t$ decreases, the TRSB window narrows and the relative-phase mode moves to the two-particle threshold: its binding energy $2\Delta_{\min}-\Omega_b$ becomes exponentially small (below $10^{-3}$~meV over most of the window), so that it cannot be distinguished from the continuum edge and is effectively damped. The $A^+$ mode remains just below the threshold.

\section*{SUPPLEMENTARY NOTE 4}
\subsection*{Robustness of the twist-angle test}

The twist-angle test of the main text (Fig.~\ref{fig:discriminant}) compares the energies of the resonances at $\Delta_{\min}+\Omega$ (inner) and $\Delta_{\max}+\Omega$ (outer) for the intrinsic collective modes with those produced by two extrinsic bosons: a fixed-energy boson with $\Omega_0=\Omega_b(\theta=45^\circ)$, and a boson tied to the intrinsic gap, $\Omega=\kappa\Delta_{\max}$, with $\kappa$ chosen such that $\Omega=\Omega_b$ at $45^\circ$. Matching the energies at $45^\circ$ is the least favourable choice, since all features then coincide at the symmetric point.The mode energies are obtained from the angular representation of the kernels given in Methods, which reproduces the values quoted in the main text.

\begin{figure}[h]
\centering
\includegraphics[width=\textwidth]{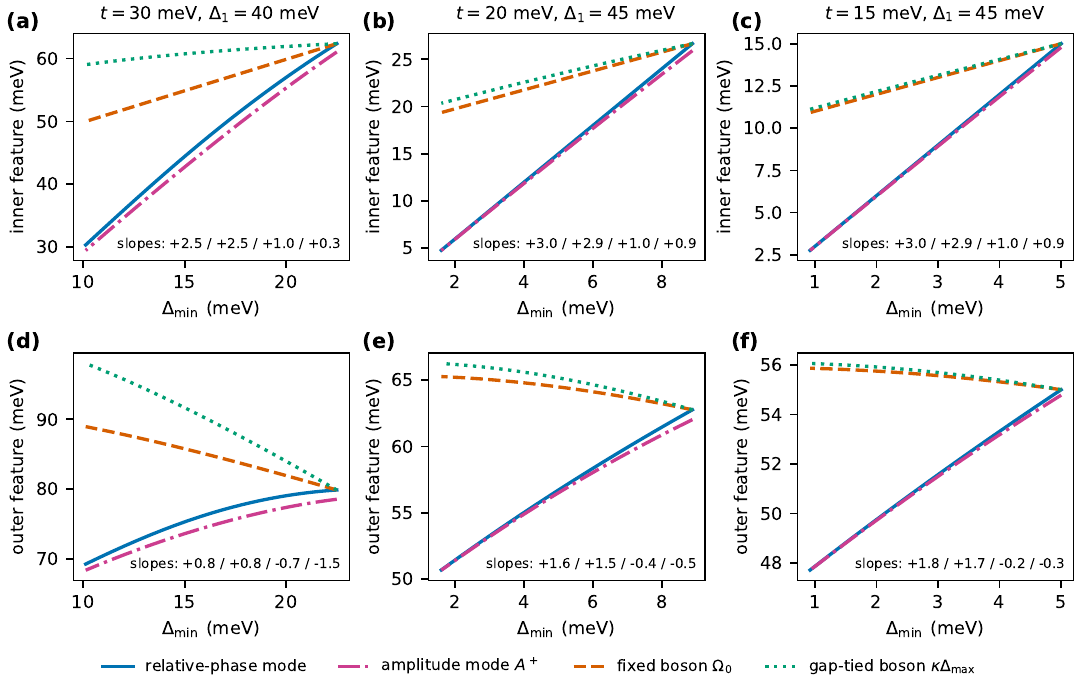}
\caption{\textbf{Twist-angle test for different interlayer tunnelling. \textbf{a}--\textbf{c}, Inner resonance energy $\Delta_{\min}+\Omega$ and \textbf{d}--\textbf{f}, outer resonance energy $\Delta_{\max}+\Omega$ against $\Delta_{\min}$ for the relative-phase mode, the amplitude mode $A^+$, a fixed-energy boson $\Omega_0=\Omega_b(45^\circ)$ and a gap-tied boson $\Omega=\kappa\Delta_{\max}$, for the three parameter sets of Supplementary Fig.~\ref{sfig:many_t}.} The fitted slopes are given in each panel in the same order and listed in Supplementary Table~\ref{stab:slopes}.}
\label{sfig:discriminant_t}
\end{figure}

Supplementary Fig.~\ref{sfig:discriminant_t} and Supplementary Table~\ref{stab:slopes} show the resonance energies against $\Delta_{\min}$ for the three values of the interlayer tunnelling of Supplementary Fig.~\ref{sfig:many_t}. In all cases the inner resonance of the collective modes moves with slope $\simeq2.5$--$3$, compared with $1$ for a fixed-energy boson and $\le1$ for a gap-tied boson, and the outer resonance moves with a positive slope, whereas for both extrinsic bosons it moves with a negative slope. The limiting values follow directly from the pinning of the modes to the pair-breaking edge: for $\Omega_i\to2\Delta_{\min}$ the inner resonance approaches $3\Delta_{\min}$ (slope $3$), and the outer one approaches $\Delta_{\max}+2\Delta_{\min}$ (slope $2+d\Delta_{\max}/d\Delta_{\min}$). Within the chiral order parameter of equation~\eqref{eq:OP_final}, $\Delta_{\max}$ changes only weakly with $\theta$ compared with $\Delta_{\min}$, so $|d\Delta_{\max}/d\Delta_{\min}|<2$ and the outer slope is positive. For $t=20$ and $15$~meV the relative-phase mode lies at the threshold (Supplementary Note~3), and the slopes of the intrinsic features are then set by the pair-breaking edge itself; the amplitude mode $A^+$ gives nearly identical slopes.

\begin{table}[h]
\centering
\caption{\textbf{Fitted slopes of the resonance energies against $\Delta_{\min}$.} Columns: relative-phase mode / amplitude mode $A^+$ / fixed-energy boson / gap-tied boson. For $t=30$~meV the fit window is $\varphi\in[\pi/6,\pi/2]$ ($31^\circ\le\theta\le45^\circ$); for $t=20$ and $15$~meV it is the TRSB window down to $\varphi=0.06\pi$.}
\label{stab:slopes}
\begin{tabular}{lcc}
\toprule
 & inner, $\Delta_{\min}+\Omega$ & outer, $\Delta_{\max}+\Omega$\\ \midrule
$t=30$ meV, $\Delta_1=40$ meV & $2.55/2.53/1.00/0.25$ & $+0.80/+0.78/-0.75/-1.50$\\
$t=20$ meV, $\Delta_1=45$ meV & $3.00/2.89/1.00/0.86$ & $+1.63/+1.53/-0.37/-0.51$\\
$t=15$ meV, $\Delta_1=45$ meV & $3.00/2.94/1.00/0.95$ & $+1.77/+1.72/-0.23/-0.28$\\ \bottomrule
\end{tabular}
\end{table}

\section*{SUPPLEMENTARY NOTE 5}
\subsection*{Interlayer charging of the relative-phase mode}
In the effective model, the relative-phase kernel, Eq.~(12) in Methods with $w_b=\sin^2 2\alpha$, consists of a static stiffness and a dynamical term $-x^2F(x)$. The low-frequency limit of the dynamical term is the phase inertia, which here is set by the quantum capacitance $C_q=e^2N$ of the top layer alone. In the bilayer, however, $\varphi$ is the phase difference between the layers, and by the Josephson relation its oscillation is accompanied by an interlayer voltage $V=\hbar\dot\varphi/2e$. The inertia is then proportional to the total capacitance $C$ of the interlayer charge transfer,
\begin{equation}
\frac1C=\frac1{C_q^{\rm top}}+\frac1{C_q^{\rm bot}}+\frac1{C_{\rm geo}},\qquad C_{\rm geo}=\frac{\varepsilon_0\varepsilon_r}{d},
\end{equation}
where $C_{\rm geo}$ is the geometric capacitance between the two CuO$_2$ planes facing the interface. We estimate the effect by rescaling the dynamical term, $x^2\to Rx^2$ with $R=C/C_q^{\rm top}$, which is exact for the low-frequency inertia and approximate near threshold. Including only the bottom-layer quantum capacitance gives $R=1/2$; with order-of-magnitude values for Bi2212 ($N\simeq1$--$3$~eV$^{-1}$ per Cu, $\varepsilon_r\simeq5$--$15$, $d\simeq1.2$--$1.5$~nm) the geometric capacitance dominates and $R\simeq1/12$--$1/200$. The reduced inertia raises the bare relative-phase frequency far above $2\Delta_{\min}$, and the binding energy $2\Delta_{\min}-\Omega_b$ vanishes rapidly: at $\theta=45^\circ$ it falls from $5$~meV in the effective model to $0.04$~meV for $R=1/2$, and is negligible for $R\lesssim1/12$. The amplitude sector does not couple to the interlayer charge at $\mathbf q=0$ in the particle--hole-symmetric model, so $\Omega_{A^+}$ is unchanged. This is an order-of-magnitude estimate; a full bilayer calculation of the coupled phase and charge fluctuations would be needed to make it quantitative, but the reduction of the inertia is large enough that the conclusion is unlikely to change.

\section*{SUPPLEMENTARY NOTE 6}
\subsection*{Second-derivative analysis for weaker interlayer coupling}
\begin{figure}[h]
  \centering
  \includegraphics[scale=.8]{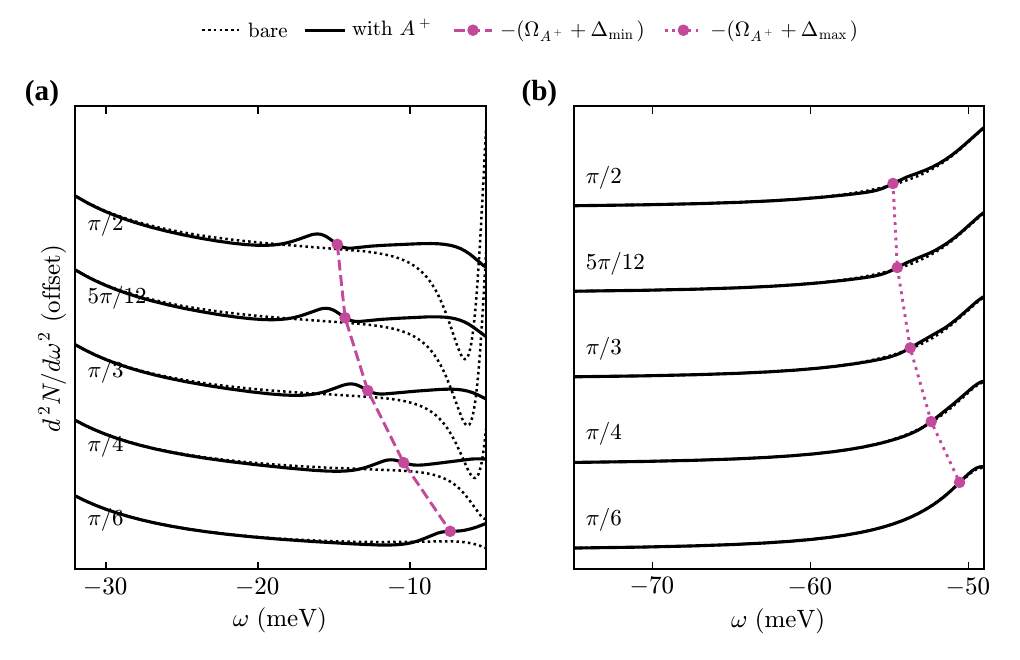}
 \caption{\textbf{Second derivative of the momentum-integrated DOS.} $d^2N/d\omega^2$ (offset) for the parameters of main text Fig.~\ref{fig:t15_combined}, near (\textbf{a}) the inner resonance $-(\Omega_{A^+}+\Delta_{\min})$ (dashed markers) and (\textbf{b}) the outer resonance $-(\Omega_{A^+}+\Delta_{\max})$ (dotted markers), for $\varphi$ between $\pi/6$ and $\pi/2$. Solid lines include the coupling to $A^+$; dotted lines show the bare spectrum.}
  \label{fig:t15_d2N}
\end{figure}

For $\Delta_1=45$~meV and $t=15$~meV the features produced by $A^+$ in the momentum-integrated DOS (main text Fig.~\ref{fig:t15_combined}a) are weak. Supplementary Fig.~\ref{fig:t15_d2N} shows the second derivative $d^2N/d\omega^2$, zoomed on the inner resonance $-(\Omega_{A^+}+\Delta_{\min})$ (a) and on the outer resonance $-(\Omega_{A^+}+\Delta_{\max})$ (b).

\end{document}